\documentclass[aps,twocolumn,prb,floatfix,showpacs,superscriptaddress]{revtex4-1}
\usepackage[columnwise]{lineno}
\usepackage{graphicx}
\usepackage{float}
\usepackage{subfigure}
\usepackage{dcolumn}
\usepackage{gensymb}
\usepackage{bm}
\usepackage{epstopdf}
\usepackage{subfigure}
\usepackage{amsmath}
\usepackage{color}
\usepackage{units}
\usepackage{scrextend}
\usepackage{amssymb}

\usepackage{color}
\usepackage[usenames,dvipsnames,svgnames,table]{xcolor}
\usepackage{hyperref}

\hypersetup{
     colorlinks   = true,
     citecolor    = blue,
     linkcolor    = blue
           }
\renewcommand{\vec}[1]{\boldsymbol{#1}}
\renewcommand{\Im}{\mathrm{Im}\,}

\newcommand{\Tr}{\mathrm{Tr}\,}

\newcommand{\euler}{\mathrm{e}}

\begin{document}
\title{Ultrafast magnetization dynamics in pure and doped Heusler and inverse Heusler alloys}
\author{R.~Chimata}
\affiliation{Argonne National Laboratory, Lemont, IL 60439, United States}
\affiliation{Department of Physics and Astronomy, Materials Theory, University Uppsala, SE-75120 Uppsala, Sweden}

\author{E.~K.~Delczeg-Czirjak}
\affiliation{Department of Physics and Astronomy, Materials Theory, University Uppsala, SE-75120 Uppsala, Sweden}

\author{J.~Chico}
\affiliation{Peter Gr\"unberg Institut and Institute for Advanced Simulation, Forschungszentrum J\"ulich \& JARA, D-52425 J\"ulich, Germany}

\author{M.~Pereiro}
\affiliation{Department of Physics and Astronomy, Materials Theory, University Uppsala, SE-75120 Uppsala, Sweden}

\author{B.~Sanyal}
\affiliation{Department of Physics and Astronomy, Materials Theory, University Uppsala, SE-75120 Uppsala, Sweden}

\author{O.~Eriksson}
\affiliation{Department of Physics and Astronomy, Materials Theory, University Uppsala, SE-75120 Uppsala, Sweden}
\affiliation{School of Science and Technology, \"Orebro University, SE-701 82 \"Orebro, Sweden}

\author{D.~Thonig}
\affiliation{Department of Physics and Astronomy, Materials Theory, University Uppsala, SE-75120 Uppsala, Sweden}
\email[Corresponding author.]{danny.thonig@physics.uu.se}

\date{\today}

\begin{abstract}
\noindent  By using a multiscale approach based on first-principles density functional theory combined with atomistic spin dynamics, we investigate the electronic structure and magnetization dynamics of an inverse Heusler and a Heusler compound and their alloys,  i. e. Mn$_{2-x}$\textit{Z}$_x$CoAl and Mn$_{2-x}$\textit{Z}$_x$VAl, where \textit{Z} = Mo, W, Os and Ru, respectively. A signature of the ferrimagnetic ordering of Mn$_{2}$CoAl and Mn$_{2}$VAl Heusler alloys is reflected in the calculated Heisenberg exchange constants. They decay very rapidly with the interatomic distance and have short range, which is a consequence of the existence of the finite gap in the minority spin band. The calculated Gilbert damping parameter of both Mn$_2$CoAl and Mn$_2$VAl is high compared to other half-metals, but interestingly in the particular case of the inverse Mn$_{2}$CoAl alloys and due to the spin-gapless semiconducting property, the damping parameters decrease with the doping concentration in clear contradiction to the general trend. Atomistic spin dynamics simulations predict ultrafast magnetisation switching in Mn$_{2}$CoAl and Mn$_{2}$VAl under the influence of an external magnetic field, starting from a threshold field of $\unit[2]{T}$. 	Our overall finding extends with Heusler and inverse Heusler alloys, the class of materials that exhibits laser induced magnetic switching.
\end{abstract}

\vspace{20mm}

\maketitle

\section{Introduction}

The field of the ultrafast magnetization dynamics has become one of the most important topics in magnetism, starting from the pioneering experiment on ferromagnetic nickel from Beaurepaire et al. \cite{Beaurepaire:1996es} in 1996. Since then, numerous experiments were carried out on 3$d$ (Fe \cite{Carpene:2008bd}, Co \cite{Cinchetti:2006cv}, Ni \cite{2007NatMa...6..740S, Rhie:2003ji}), 4$f$ (Tb and Gd \cite{Wietstruk:2011ex}) ferromagnets, as well as on several alloys (GdFeCo \cite{Stanciu:2006gm,Vahaplar:2009fpa,Stanciu:2007fy,Steil:2011dt,Ostler:2012hx,Mentink:2012ga,Chimata:2015bea}, TbCo \cite{Alebrand:2014fp}, CoPt \cite{Bigot:2009bu}) and half metallic systems (CrO$_2$ \cite{Muller:2008ip}, Co$_{2}$Cr$_{0.6}$Fe$_{0.4}$Al \cite{Wustenberg:2011co}, Co$_{2}$FeSi, Co$_2$MnGe, Co$_2$FeAl \cite{Mann:2012bra}, and Co$_{2}$Fe$_x$Mn$_{1-x}$Si \cite{Steil:2010fi}  and Co$_{2}$MnSi \cite{Steil:2010fi, Liu:2010cr}) aiming to find faster ways of manipulating spins in a controllable way, opening a new field in the advanced information/data storage and data processing technologies.

Experimental observations revealed that the characteristic demagnetization times of 3$d$ elements are within the $\unit[100]{fs}$ time scale, much faster than that of the 4$f$-ferromagnets, which show more complex behavior involving a two-step demagnetization process in $\unit[10]{ps}$ time scale. Surprisingly, recent pump-probe experiments on half-metallic Heusler alloys measured distinguished and typically larger all-optical switching times when compared to 3$d$-ferromagnets \cite{Steil:2010fi,Vahaplar:2009fpa}. In these materials, one of the spin channels is completely or partially unoccupied around the Fermi energy, consecutively the magneto optical excitations from one channel to another channel are forbidden.

Attempts to understand the momentum transfer between the electrons, spins and phonons after a short laser pulse have opened a new debate in the field. Several quantitative models had been proposed to describe the mechanism of the ultrafast demagnetization such as the microscopic three-temperature model \cite{Koopmans:2009ds}, stochastic atomistic descriptions \cite{Radu:2011kr}, models using the stochastic Landau-Lifshitz-Bloch equation \cite{Kazantseva:2008bq,Atxitia:2007cq} and models suggesting the presence of diffusive or superdiffusive spin currents \cite{Battiato:2010br,Melnikov:2011ep,Wietstruk:2011ex,Carva:2011dp,Essert:2011iq}. The first three models relate the spin-scattering to the Gilbert damping parameter, $\alpha$, that describes the energy dissipation in a magnetic system via elementary spin-flip processes \cite{Atxitia:2011gn, Fahnle:2011eg}. Here, we combine the {\it ab initio} description of the magnetic exchange interaction and Gilbert damping \cite{Gilmore:2007evb,Brataas:2008eqb,Ebert:2011gx} parameter with the Landau-Lifshitz-Gilbert equation to investigate the demagnetization process in half-metallic ferrimagnetic Heusler and inverse Heusler alloys.

\begin{figure}[ht]
\centering
\subfigure{\includegraphics[scale=1.2]{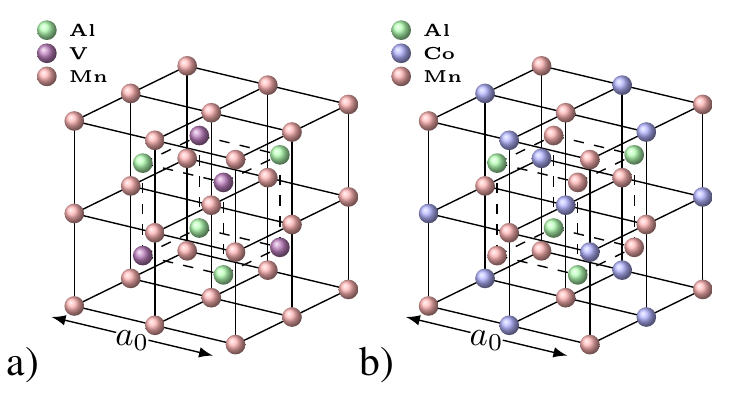}}
\caption{(Color online) Schematic crystal structures of a) the Heusler alloy Mn$_2$VAl and b) the inverse Heusler alloy Mn$_2$CoAl. Different atom types are represented by different colours. Solid and dashed lines indicate the bond between the atoms and are added to guide the eye. The lattice constant $a_0$ is also indicated.}
\label{fig:1}
\end{figure}

Heusler and inverse Heusler alloys are defined as ternary intermetallic compounds with a composition of $X_2YT$ (cf. Fig.~\ref{fig:1}). Heusler alloys crystallize in the L2$_1$ structure (space group Fm$\bar{3}$m, 225), with the 4$a$ (0, 0, 0), 4$b$ ($\frac{1}{2}$, $\frac{1}{2}$, $\frac{1}{2}$) and 8$c$ ($\frac{1}{4}$, $\frac{1}{4}$, $\frac{1}{4}$) Wyckoff positions. $X$ and $Y$ are transition metals occupying the 8$c$ and 4$a$ positions, respectively, and $T$ is a main group III, IV or V element sitting in the 4$b$ position. Inverse Heusler alloys adopt the Hg$_2$CuTi prototype structure (space group F$\bar{4}3$m, 216), with 4$a$ (0,0,0), 4$b$ ( $\frac{1}{2}$, $\frac{1}{2}$, $\frac{1}{2}$), 4$c$ ( $\frac{1}{4}$, $\frac{1}{4}$, $\frac{1}{4}$ ) and 4$d$ ($\frac{3}{4}$, $\frac{3}{4}$, $\frac{3}{4}$) positions. In this case, $X$ and $Y$ are transition metals, $X$ occupying the 4$a$ and 4$d$ positions while $Y$ is the 4$c$ position. The main element T sits in the 4$b$ position. Both structures may be regarded as a cubic unit cell,  which consists of four interpenetrating fcc sublattices. There are four atoms in the diagonal of the cube following the $X$-$Y$-$X$-$T$ sequence for Heusler alloys and $X$-$X$-$Y$-$T$ for the inverse Heuslers.

Here, we study the demagnetization dynamics of a Heusler and an inverse Heusler compound and their alloys, i.e. Mn$_{2-x}Z_x$VAl and Mn$_{2-x}Z_x$CoAl, where $\textit{Z}$ = Mo, W, Os and Ru. Mn$_2$VAl is a well known half-metallic ferrimagnetic Heusler compound \cite{Itoh:1983hy, Yutaka:2013ki, Jiang:2001kd, Shoji:2013ge, Ozdogan:2006jc,Sasioglu:2005ex} where the minority spin channel is the conducting one \cite{Weht:1999gq}. Mn$_2$CoAl adopts the inverse Heusler structure \cite{Liu:2008dt} and it is predicted \cite{Liu:2008dt} and confirmed \cite{Ouardi:2013ko} to be a spin gapless magnetic semiconductor. These peculiarities of the band structure are reflected in the Gilbert damping parameter and affect the magnetisation dynamics under the influence of a laser pulse, as will be described below.

The paper is divided as follows: In Section~\ref{sec:method} we introduce our numerical technique to study materials properties and magnetization dynamics in Heusler alloys. Electronic and magnetic properties of the parent Heusler alloys Mn$_2$CoAl and Mn$_2$VAl as well as doping of these materials with Os, Ru, W, and Mo is discussed in Section~\ref{sec:elst}. Demagnetisation studies of these alloys caused by a femtosecond laser are described in Section~\ref{sec:dynamics}. Finally, the article concludes in Section~\ref{conclusion} with an outlook.

\section{METHODS}
\label{sec:method}
\subsection{Electronic structure calculation}
The electronic and magnetic properties of the studied materials are obtained from first principle calculations by applying full-relativistic multiple scattering theory as formulated in the Korringa-Kohn-Rostocker (KKR) approach \cite{Zabloudil:2005dJ}. This method is implemented in the SPR-KKR package\cite{Ebert:2011di,Ebert:PzMUDULG}. Solving the Dirac equation, relativistic effects are fully accounted for, especially the spin-orbit interaction which is essential for heavy elements such as the here considered dopants Os, W, Ru, and Mo. The potentials are treated by the atomic sphere approximation (ASA) and obtained by self-consistently solving the Kohn-Sham density functional theory (DFT) equation within the local density (LDA) or generalized gradient approximation (PBE) as devised by Perdew, Burke and Ernzerhof \cite{Gross:2013jq,Perdew:1996iq}. Note that we applied the PBE functional if not further specified. The irreducible Brillouin zone is sampled by $\approx 500$ k-points. To describe substitutional disorder in the sub-lattices of the alloys we make use of the coherent potential approximation (CPA)\cite{Gyorffy:1972df}. The spin-polarized scalar relativistic full-potential (SR-FP) mode\cite{1998PhRvB..5810236H} of the KKR approach is used to calculate the total energies as a function of volume [$E(V)$], which gives an estimate of the lattice constant $a_0$.


\subsection{Calculation of Heisenberg exchange and Gilbert damping}

The angular momentum transfer in terms of Heisenberg exchange interactions $J_{ij}$  and energy dissipation related to the Gilbert damping parameter $\alpha$ is determined by an ab-initio method with the aim to address the magnetic ground state and also the dynamical properties by using the Landau-Lifshitz-Gilbert equation. The interatomic exchange interactions, $J_{ij}$, were calculated via the Liechtenstein-Katsnelson-Antropov-Gubanov (LKAG) formalism \cite{Liechtenstein:1984fj}

\begin{align}
J_{ij}=\frac{1}{\pi}\int_{-\infty}^{\varepsilon_F} \Im \Tr \left(\Delta_{i}\tau^\uparrow_{ij}\Delta_{j}\tau^\downarrow_{ji}\right)\mathrm{d}\varepsilon	 .
\label{eq:jij}
\end{align}
where $\Delta_i = t_{i,\uparrow}^{-1}-t_{i,\downarrow}^{-1}$ is the spin-resolved difference of the single-site scattering matrix $t_i$ at site $i$ and $\tau_{ij}$ is the scattering path operator, describing the propagation of the electrons between two sites $i$ and $j$. The Fermi energy is denoted by  $\varepsilon_F$. Note that in CPA, the multiple scattering matrix is replaced by the scattering properties of the effective medium $\hat{\tau}_{i\mu,j\nu}=X_{i\mu}\tau_{ij}^{CPA}X_{j\nu}$ constructed from a defect of type $\mu, \nu$ at site $i,j$, respectively. The defects are taken into account by the defect matrix $X_{i\mu}$. From the calculated exchange interactions, it is possible to obtain the spin wave stiffness, $D$, which is expressed as:\cite{Pajda:2001ix}
\begin{align}
D=\lim_{\eta\rightarrow 0}\frac{2}{3}\sum_{j}\euler^{-\eta \frac{\left|\vec{r}_{0j}\right|}{a_0}}J_{0j}\left|\vec{r}_{ij}\right|^2	
\end{align}
by using super cell calculation with random configurations of the dopants in $12$ ensembles and starting from a reference site $i=0$. The distance between site $i$ and $j$ is given by $\vec{r}_{ij}$ and the parameter $\eta$ is introduced to guarantee  convergence within a Pade interpolation approximation.  

The Gilbert damping  parameter is identified on the basis of the linear response theory  \cite{Ebert:2011gx} by means of the multiple scattering formalism \cite{Mankovsky:2013ii}. The diagonal elements $\mu=x,y,\textit{Z}$ of the Gilbert damping tensor can be written as \cite{Ebert:2011gx}:

\begin{eqnarray}
\alpha^{\mu\mu} = \frac{g}{\pi m_{tot}} \sum_{j } \Tr
\left\langle \mathcal{T}^{\mu}_0 \,
 \tilde{\tau}_{0j}\,
\mathcal{T}^{\mu}_{j} \,
 \tilde{\tau}_{j0} \right\rangle_{c} \; ,
\label{alpha_MST}
\end{eqnarray}
where the effective g-factor  $g = 2(1 + {m_{orb}}/{m_{spin}})$ and  total magnetic moment $ m_{tot} = m_{spin} + m_{orb} $ are given by the spin and orbital moments, $m_{spin}$ and $m_{orb}$, respectively, ascribed to a unit cell. Equation~\ (\ref{alpha_MST}) gives $\alpha^{\mu\mu}$ for the atomic cell at lattice site $0$ and implies a summation over contributions from all sites indexed by $j$, including $j=0$. Moreover, $\tilde{\tau}_{ij}$ is related to the imaginary part of the multiple scattering operator that is evaluated only at the Fermi energy $\varepsilon_F$. Finally, $\mathcal{T}^{\mu}_i$ represents the matrix elements of the torque operator $\hat{\mathcal{T}}^\mu=\beta\sigma^\mu B_{xc}(\vec{r})$. The notation $\langle\hdots\rangle_c$ represents the configurational average, including vertex corrections\cite{Ebert:2011gx} derived by Butler \cite{Butler:1985by} and accounting for finite temperature using the alloy analogy model within CPA\cite{Ebert:2015kxa}.

\subsection{Atomistic spin dynamics}

The evolution of atomistic spins in a thermal bath is described by the Landau-Lifshitz-Gilbert (LLG) equation~\cite{Antropov:1996td,Eriksson:2016uw}, where the dynamics of a magnetic moment is expressed in terms of precession and damping:

\begin{equation}
\label{llge}
\begin{split}
\frac{d{\vec{m}}_{i}(t)}{dt} &= -\frac{\gamma}{(1+\alpha^2)} \biggl( { \vec{ m}}_{i}(t)\times \vec{B}_{i}(t) \\  &+  \frac{\alpha}{{\textit m_i}} {\vec{m}}_{i}(t)\times ({\vec{m}}_{i}(t)\times \vec{ B}_{i}(t))\biggr).
\end{split}
\end{equation}

Here $\gamma$ is the gyromagnetic ratio, $\alpha$ represents the dimensionless Gilbert damping constant, and ${\vec{m}}_i=m_i\vec{e}_i$ is an individual atomic moment on site $i$. The effective magnetic field is given by $\vec{B}_{i}=-\frac{\partial {\cal H} }{\partial \vec{m}_{i}}+{\vec{b} }_i$, where $\mathcal{H} =-\sum_{i\neq j}J_{ij}\vec{e}_{i}\cdot\vec{e}_{j}$ and $\vec{b}_{i}$ is an stochastic field. The latter describes white noise ($\langle \vec{b}_i(t)\cdot\vec{b}_j(t^\prime)\rangle=2D\delta_{ij}\delta(t-t^\prime)$), where the fluctuation width is $D=\nicefrac{\alpha k_B T_s}{\gamma m}$. Thus, the spin temperature $T_s$ directly passes into LLG equation via the stochastic magnetic field $\vec{b}_{i}$ and is obtained from solving the two-temperature (2T) model\cite{Bovensiepen:2007gc}. The analytical expression of this two temperature model reads,

\begin{eqnarray}
\label{tt}
T_{s}  & = & T_{0} + \\  \nonumber && (T_{P}-T_{0})\times(1- \exp^{(-t/\tau_{\rm initial})})\times \exp^{(-t/\tau_{\rm final})} + \\  \nonumber && (T_{F}-T_{0})\times(1-\exp^{(-t/\tau_{\rm final})})
\end{eqnarray}
where $T_{\rm 0}$ is the initial temperature of the system, $T_{\rm P}$ is the peak temperature after the laser pulse is applied and $T_{\rm F}$ is the final temperature. Both the initial and final temperature are set to $\unit[300]{K}$, where the peak temperature is a parameter in the simulations. The time-dependent parameters $\tau_{\rm initial}$ and $\tau_{\rm final}$ are exponential parameters, fixed by $\tau_{\rm initial}=\unit[10]{fs}$ and $\tau_{\rm final}=\unit[20]{ps}$ from Ref.~[\onlinecite{Chimata:2012kv}]. Note that both relaxation times are materials specific and $k_{\rm B}$ is the Boltzmann constant.

\section{RESULTS AND DISCUSSION}
\label{sec:results}

This current section is divided in five parts. In the first and second part we discuss the electronic structure and the magnetic moments, respectively, of pure and doped Heusler and inverse Heusler materials based on DFT-optimized lattice constants. The third part deals with the Heisenberg interaction, spin wave stiffness, as well as the ordering temperature. The Gilbert damping is discussed in part four. The last part focuses on the demagnetisation and reliable switching in Heusler materials based on the LLG equation. 
 
\subsection{Electronic structure calculations}
\label{sec:elst}

Lattice parameters are estimated from total energy calculations, compared to  Refs.~[\onlinecite{Ozdogan:2006jc}] and [\onlinecite{Jakobsson:2015dt}], and listed in Table~\ref{tab:1}. For undoped Mn$_2$CoAl and Mn$_2$VAl, we improved the theoretically predicted values used in Ref.~[\onlinecite{Jakobsson:2015dt}] by $10\%$ and they are closer to the experimentally measured lattice constant. The improvement comes from taking into account the full-potential, which is known to improve lattice constants \cite{Motizuki:2009vs}. By doping Mn with 4d and 5d metals Mo, Ru, W, and Os, we observe an expected increase of the lattice constant with the concentration of the dopants, since the atomic radius of the dopant is larger than the one of Mn. For Mn$_2$VAl, the increase of the lattice constant is substantially bigger ($\approx 1\%$ for $x=1\%$ doping) than for Mn$_2$CoAl ($\approx 0.1\%$ for $x=1\%$ doping).

\begin{figure}[t]
\centering
\includegraphics[width=1.15\columnwidth]{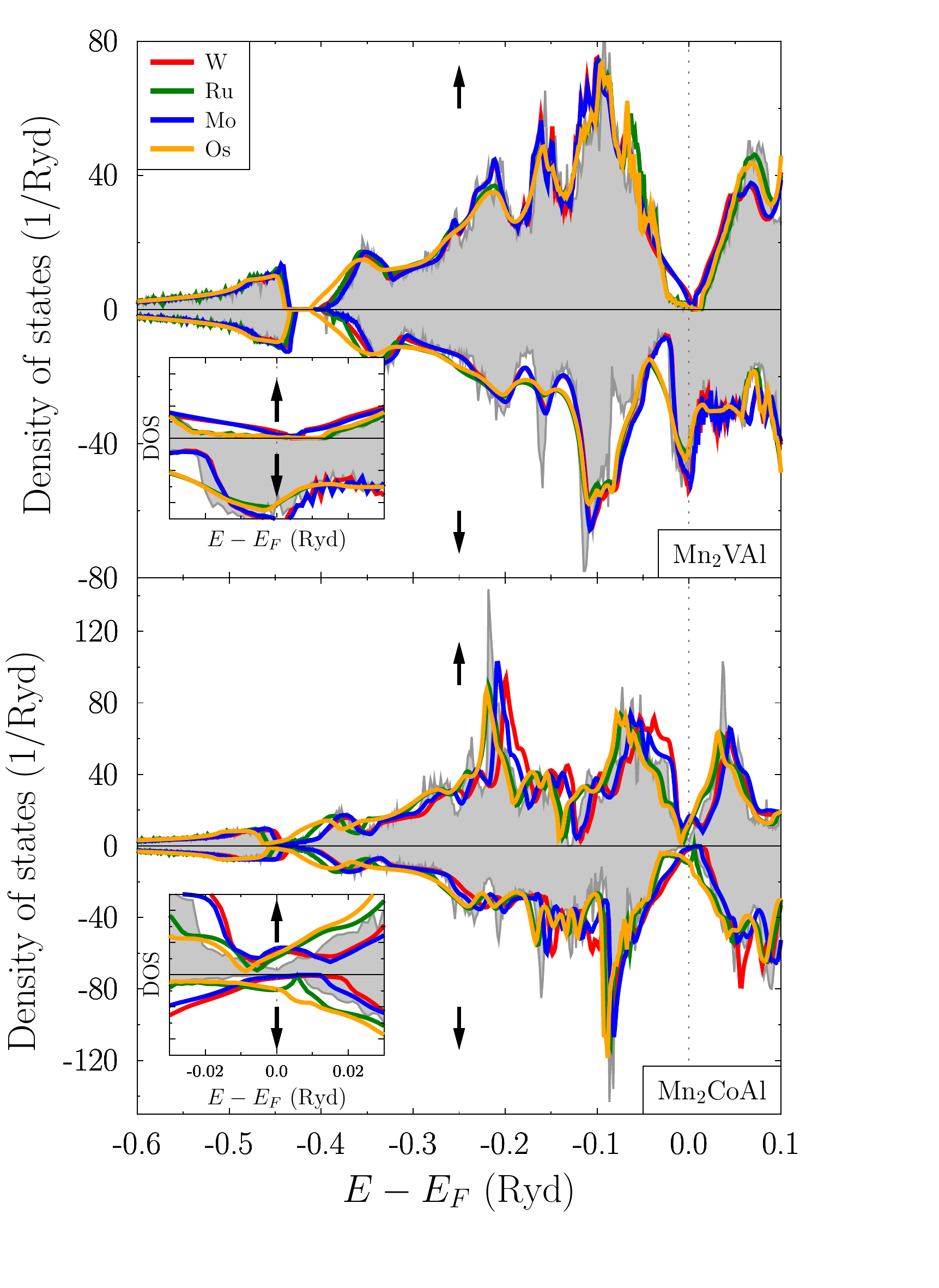}
\caption{(Color online) Density of states (DOS) for Mn$_2$CoAl (lower panel) and Mn$_2$VAl (upper panel) without doping(gray background) and with doping of W (red lines), Ru (blue lines), Mo (green lines), and Os (orange lines). Positive (negative) DOS values correspond to the majority-(minority-) spin electrons and are indicated by bold up-(down-) arrays. The inset is a magnification of the DOS around the Fermi level.}
\label{fig:2}
\end{figure}

Thermal switching within our classical atomistic model is completely determined by the Heisenberg exchange and the Gilbert damping of the system \cite{Chimata:2017iu}, which are in turn identified by the scattering-path matrices and the single-site scattering matrices of the Kohn-Sham problem in Eqs.~\eqref{eq:jij} and \eqref{alpha_MST}. Hence, we first have to address the electronic structure by means of the density of states (DOS; Fig.~\ref{fig:2}). The here studied inverse Heusler Mn$_2$CoAl is known to be a spin gapless semiconductor, where an almost zero-width energy gap at the Fermi level exists in the majority-spin channel (the majority states are plotted with positive values and the minority spin states with negative values) but a regular energy gap occurs in the minority spin-channel (see inset in the bottom panel of Fig.~\ref{fig:2}). This was already reported, for example, in Ref.~[\onlinecite{Jakobsson:2015dt}]. The density of states and, consequently, the gap are sensitive to the applied exchange correlation functional. Using local density approximation (LDA), states are shifted up in energy (not shown here) compared to the PBE by about $\unit[10]{meV}$ and, consequently, no gap at the Fermi energy is observed. Note that the offset of the energy from the real axis in Fig. \ref{fig:2} (the spectral width of the electron bands) is small and about $\unit[1]{meV}$, which causes sharp features in the DOS. A finite spectral width also gives rise to an overlap of the states around the Fermi level and `hide' the zero-width energy gap; a finite density of states at $\varepsilon_F$ is observed. Bands that cross the Fermi level, are mainly allocated to Mn$^1$ and Co (band structure is not shown here, but it can be found elsewhere \cite{Liu:2008dt}). Note that the superscripts $1$ and $2$ between the two Mn atoms. In contrast to Ref.~[\onlinecite{Jakobsson:2015dt}], the Fermi energy is not located at the centre of the minority band gap, which will affect the coupling between the collective and single-electron excitations, i.e. the exchange interactions.

The chemical compound Mn$_2$VAl, however, is half-metallic (cf. Fig.~\ref{fig:2}) with a gap in the majority spin-channel. The width of the majority band gap ($\unit[0.7]{eV}$) is bigger than the minority spin gap in Mn$_2$CoAl ($\unit[0.4]{eV}$), which significantly affects the magnetic properties. In the minority spin channel and at the Fermi energy Mn projected states cause a strong peak in the DOS that hybridize with V atoms. States above the Fermi energy are dominated by the d-states of V atoms.


The spin-gapless semiconducting or half-metallic behaviour in  Mn$_{2-x}$\textit{Z}$_{x}$CoAl and Mn$_{2-x}$\textit{Z}$_{x}$VAl is destroyed by replacing some of the Mn atoms with heavy metals, \textit{Z} = Mo, W, Os, Ru of a given concentration $x=0.05$ and $0.1$. Comparing total energies (not shown here) allows us to conclude that for the inverse Heusler Mn$_{2-x}$\textit{Z}$_{x}$CoAl doping at both Mn-sites (Mn$^1$-Mn$^2$) has the lowest energy. We obtained a maximal energy difference of $\Delta E \approx \unit[40]{meV}$ when doping at Mn$^1$-\textit{Y}, Mn$^2$-\textit{Y}, or \textit{Y} with $1\%$ of the dopants W, Ru, Mo, Os. There is no major variation found in $\Delta E$ between the different dopands. Note that we used here the same lattice constant as shown in Table~\ref{tab:1}, but in principle it will vary when doping at Mn$^1$-Mn$^2$, Mn$^1$-\textit{Y}, Mn$^2$-\textit{Y}, or \textit{Y}. However, Mn$_{2-x}$\textit{Z}$_{x}$VAl has the lowest energy when doping only the V atom, but to treat both material on the same footing, we consider also Mn$_{2-x}$\textit{Z}$_{x}$VAl to be doped at the Mn$^1$-Mn$^2$ atoms.  

In the case of Mn$_2$CoAl, W and Mo generate states at the spin-gap majority states at the Fermi level, where on the other hand the gap in the minority spin channel survives. In terms of a rigid band model, W- and Mo-doping decreases the Fermi energy, which relocates the DOS to higher energies. The dopants Os and Ru have one electron more than Mn in the valence band and, consequently, affect the density of states in the opposite way: Minority states are added and become occupied. The Fermi energy increases, which shifts the density of states to smaller energies. For Mn$_2$VAl, doping with Ru and Os preserves the half-metallic behaviour; it add states below the Fermi energy and typically at the energy $\varepsilon=\unit[-0.025]{Ryd}$. Doping with Mo and W reduces the width of the band-gap and shifts it above the Fermi energy. Related to the alloying, the density of states smears out in the whole energy range. 

\subsection{Magnetic moments}
\label{sec:magmom}

\begin{figure}[ht]
\centering
\includegraphics[width=0.8\columnwidth]{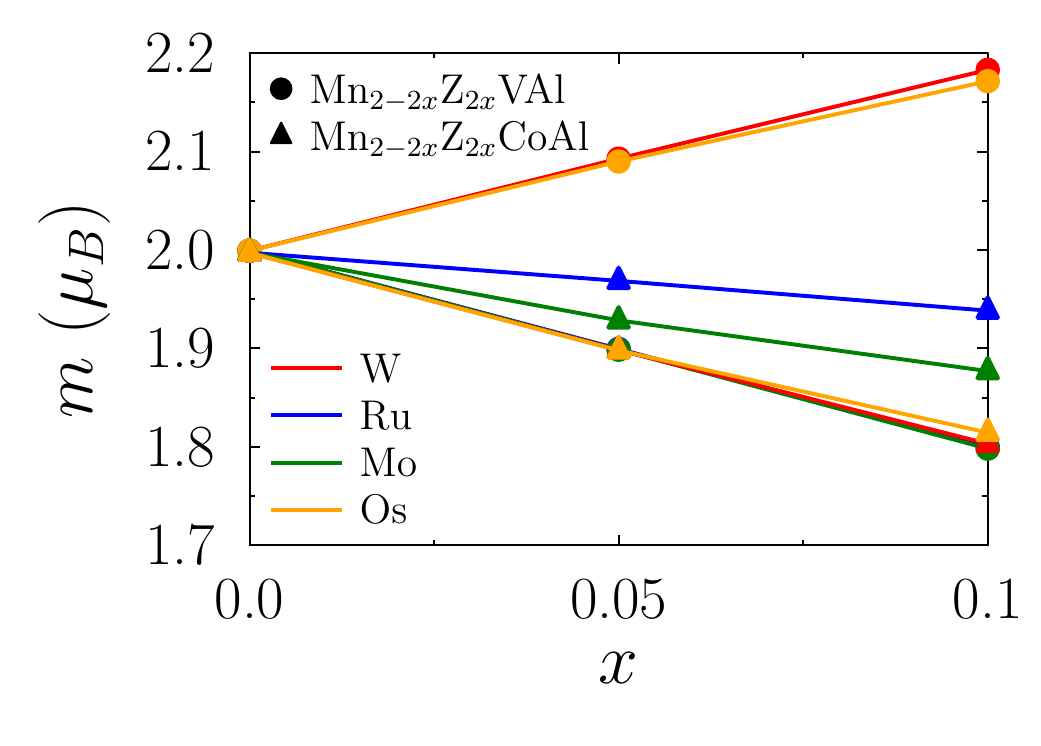}
\caption{(Color online) Total magnetic moments of Mn$_{2-x}$\textit{Z}$_{x}$CoAl (triangles) and Mn$_{2-x}$\textit{Z}$_{x}$VAl (circles) as a function of dopant concentration $x$. The symbol Z represents Mo (green lines and symboles), Os (orange lines and symbols), Ru (blue lines and symbols), and W (red lines and symbols).}
\label{fig:3}
\end{figure}

The exchange splitting in the DOS and, consequently, the total magnetic moment is affected by doping  (see Fig.~\ref{fig:3}). Both Heusler materials are ferrimagnetic. An antiferromagnetic coupling between the Mn atoms was observed for the inverse Heusler alloy Mn$_2$CoAl (cf. Table~\ref{tab:1}), caused by the inequivalence of the two Mn atoms. These results are in good agreement with experiments \cite{Liu:2008dt,Ouardi:2013ko} and existing theoretical predictions \cite{Jakobsson:2015dt,Meinert:2011hn}. According to the Bethe-Slater curve \cite{Jiles:2015uu}, transition-metal atoms such as Mn tend to have an antiferromagnetic spin moment when they are close to each other. In Mn$_{2-x}$\textit{Z}$_{x}$VAl, the Mn atoms are equivalent and, thus, have the same magnetic moment that couple ferromagnetically. The V atom, however, is antiferromagnetic with respect to the Mn atoms and has a strong induced magnetic moment of $\unit[0.91]{\mu_B}$. Opposite to the total magnetic moment, the size of the element resolved magnetic moments is sensitive to the lattice constant of the system and moments can vary up to $\unit[13]{\%}$, which was also found in Ref.~\onlinecite{Jakobsson:2015dt}.  

\begin{table*}
	\begin{tabular}{c c c c c c c }
		\hline
		\hline
		Compound & $a_0$ (\AA) & $m_{[Mn^1]}$ & $m_{[Z^1]}$ & $m_{[Mn^2]}$ & $m_{[Z^2]}$ & $m_{[Y]}$ \\
		\hline
		Mn$_2$CoAl & $5.73$ [~\onlinecite{Jakobsson:2015dt}] & $-1.64$ &         & $2.77$ &        & $0.93$ \\ 
		Mn$_{1.8}$W$_{0.2}$CoAl  &             & $-1.52$ & $-0.52$ & $2.75$ & $0.26$ & $0.78$  \\ 
		Mn$_{1.8}$Ru$_{0.2}$CoAl &             & $-1.62$ & $-0.10$ & $2.76$ & $0.06$ & $0.91$  \\ 
		Mn$_{1.8}$Mo$_{0.2}$CoAl &             & $-1.53$ & $-0.56$ & $2.75$ & $0.33$ & $0.78$  \\ 
		Mn$_{1.8}$Os$_{0.2}$CoAl &             & $-1.56$ & $-0.12$ & $2.76$ & $0.18$ & $0.92$  \\ 
		
		Mn$_2$VAl  & $5.69$ [~\onlinecite{Ozdogan:2006jc}]& $1.32$  &         & $1.32$ &        & $-0.66$ \\ 
		Mn$_{1.8}$W$_{0.2}$VAl   &             & $1.32$  & $0.19$  & $1.32$ & $0.19$ & $-0.57$ \\ 
		Mn$_{1.8}$Ru$_{0.2}$VAl  &             & $1.31$  & $0.08$  & $1.31$ & $0.08$ & $-0.61$ \\ 
		Mn$_{1.8}$Mo$_{0.2}$VAl  &             & $1.36$  & $0.25$  & $1.36$ & $0.25$ & $-0.58$ \\ 
		Mn$_{1.8}$Os$_{0.2}$VAl  &             & $1.31$  & $0.09$  & $1.31$ & $0.09$ & $-0.60$ \\ 
		\hline
		\hline	
		Mn$_2$CoAl                 & $5.79$ [$5.84$ exp]  & $-1.81$ &         & $2.91$ &        & $0.96$\\ 
		Mn$_{1.8}$W$_{0.2}$CoAl  & $5.79$  & $-1.37$ & $-0.46$ & $2.62$ & $0.22$ & $0.77$  \\ 
		Mn$_{1.8}$Ru$_{0.2}$CoAl & $5.79$  & $-1.81$ & $-0.10$ & $2.90$ & $0.07$ & $0.96$  \\ 
		Mn$_{1.8}$Mo$_{0.2}$CoAl & $5.79$  & $-1.71$ & $-0.60$ & $2.89$ & $0.39$ & $0.82$  \\ 
		Mn$_{1.8}$Os$_{0.2}$CoAl & $5.80$  & $-1.80$ & $-0.12$ & $2.92$ & $0.19$ & $0.98$  \\ 
		
		Mn$_2$VAl                  & $5.84$ [$5.88$  exp] & $1.47$  &         & $1.47$ &        & $-0.91$\\ 
		Mn$_{1.8}$W$_{0.2}$VAl   & $5.92$    & $1.73$  & $0.37$  & $1.73$ & $0.37$ & $-0.99$ \\ 
		Mn$_{1.8}$Ru$_{0.2}$VAl  & $5.86$    & $1.50$  & $0.05$  & $1.50$ & $0.05$ & $-0.89$ \\ 
		Mn$_{1.8}$Mo$_{0.2}$VAl  & $5.91$    & $1.72$  & $0.45$  & $1.72$ & $0.45$ & $-0.98$ \\ 
		Mn$_{1.8}$Os$_{0.2}$VAl  & $5.92$   & $1.52$  & $0.06$  & $1.52$ & $0.06$ & $-0.91$ \\ 
		\hline		
		\hline
	\end{tabular}
	\caption{Lattice constant and atom resolved magnetic moments (in $\mu_B$) of the host Mn$_2$CoAl and Mn$_2$VAl . The upper panel shows results for a fixed lattice constant obtained from literature, where the lower panel is for lattice constants calculated from total energy minimization. The superscripts $1$ and $2$ distinguish between the two Mn atoms. The symbol $Y$ represents either Co or V. The magnetic moment of Al is negligibly small. }
	\label{tab:1}
\end{table*}

The size but not the sign of the elemental magnetic moments changes by doping the Heusler materials with 4d and 5d heavy metals, and, thus, also the total magnetic moment. Typically, the induced magnetic moments of dopants are parallel to the magnetic moment of Mn atoms and they become larger if the magnetic moment of the Mn atom is smaller. In the case of  Mn$_2$CoAl, the dopants W, Ru, Mo, and Os cause a decay of the total magnetic moment of about $\unit[0.1 - 0.2]{\mu_B}$ for $x=1\%$, while in the case of  Mn$_2$VAl, only the dopants  Ru and Mo decrease the magnetic moment.  This is caused by a significant change of the Mn magnetic moments of about $\Delta m\approx\unit[0.1-0.2]{\mu_B}$, but also for  Co atoms the moment variation is about  $\Delta m\approx\unit[0.2]{\mu_B}$.

\subsection{Heisenberg exchange parameter and Curie temperatures}

\begin{figure}[ht]
\centering
\includegraphics[width=0.7\columnwidth]{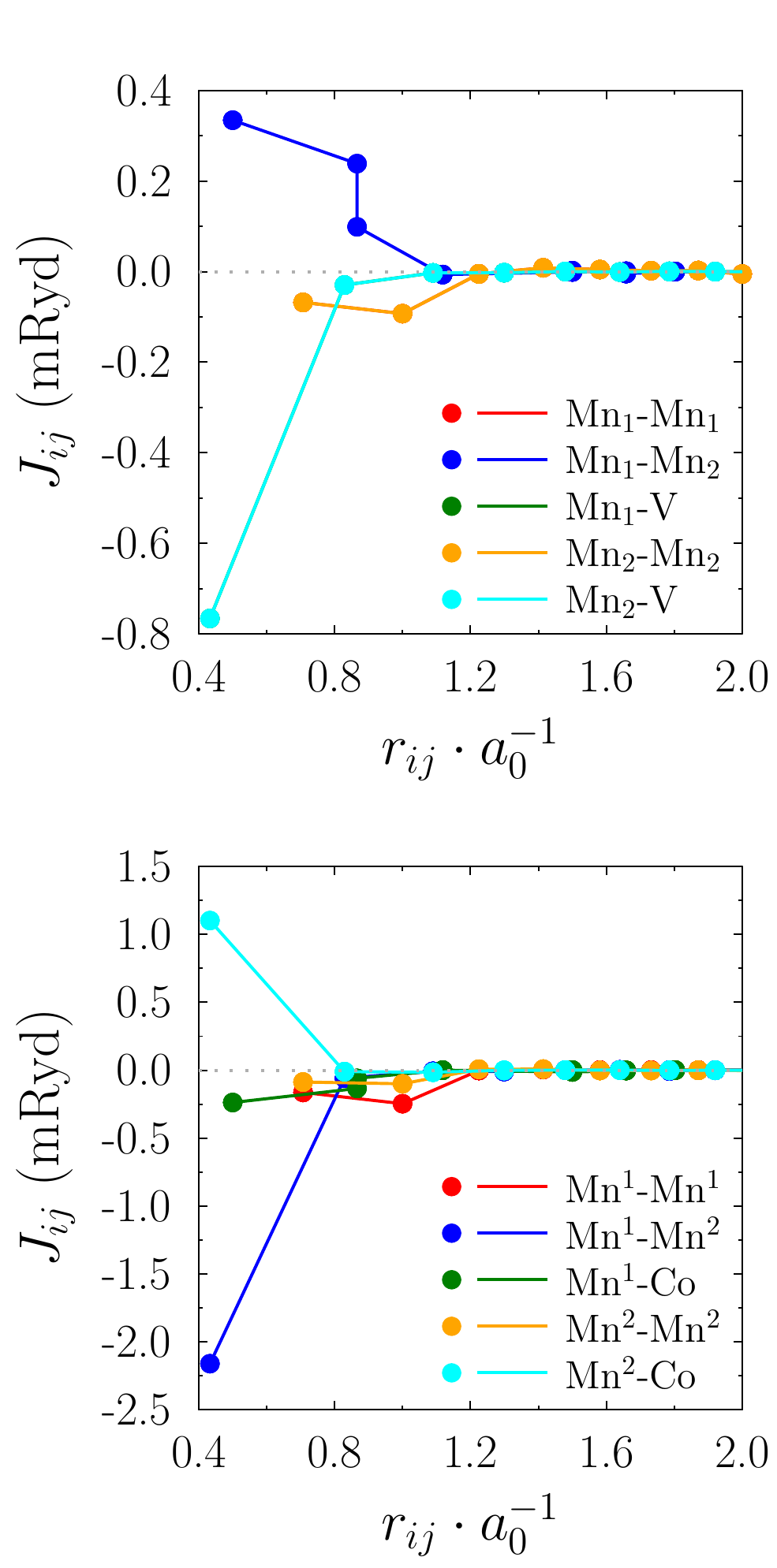}
\caption{(Color online) Intersublattice Heisenberg exchange parameter as a function of renormalized interatomic distance for a) Mn$_2$VAl and b) Mn$_2$CoAl. Different colours represents the coupling between Mn$^1$-Mn$^1$ (red dotes), Mn$^1$-Mn$^2$ (blue dotes), Mn$^1$-Co or Mn$^1$-V (green dotes), Mn$^2$-Mn$^2$ (orange dotes) and Mn$^2$-Co or Mn$^2$-V (cyan dotes).}
\label{fig:4}
\end{figure}

Based on our electronic structure analysis in the Section~\ref{sec:magmom}, we calculated the Heisenberg exchange parameter $J_{ij}$ (see Fig.~\ref{fig:4}). The already revealed ferrimagnetic behaviour is reflected also in the exchange constants $J$. The magnetic exchange parameters decay very rapidly with the interatomic distance, $r_{ij}$, which is ascribed to the existence of the finite spin gap in the minority-channel \cite{Pajda:2001ix,Rusz:2006jy}. Our results for Mn$_2$CoAl are similar  to the ones already reported in Refs.~[\onlinecite{Meinert:2011hn},\onlinecite{Jakobsson:2015dt}]. Note the factor of $2$ in Ref.~[\onlinecite{Jakobsson:2015dt}] may be caused by a different double-counting convention of the Heisenberg Hamiltonian. For the compound Mn$_2$CoAl, the antiferromagnetic interaction between Mn$^1$ and Mn$^2$ dominates the ferrimagnetism, whereas the Mn$^2$-Co interatomic exchange interaction is ferromagnetic. In Mn$_2$VAl, the situation is the opposite: the Mn to V interaction is dominating and antiferromagnetic, where only the Mn$^1$-Mn$^2$ contributes with a ferromagnetic coupling but with half the strength of the Mn-V interaction. The coupling between equivalent Mn atoms in Mn$_2$VAl (Mn$^1$-Mn$^1$ and Mn$^2$-Mn$^2$) is small and negligible. The calculated interactions depend to some extent on the details of the calculations. In particular, the $J^{\text{Mn-Co}}$ and $J^{\text{Mn-V}}$ interactions depend strongly on the applied exchange-correlation functional, but also on the lattice constant of the system. Notice that for $J^{\text{Mn-Co}}$ and $J^{\text{Mn-V}}$ in LDA we obtain twice the size of the $J$'s from PBE (not shown here). The other couplings (e.g. $J^{\text{Mn-Al}}$, $J^{\text{Co-Al}}$, $J^{\text{V-Al}}$) turned out to be negligible, primarily caused by a vanishing magnetic moment on the Al atom. 

\begin{figure}[ht]
\centering
\includegraphics[width=1.0\columnwidth]{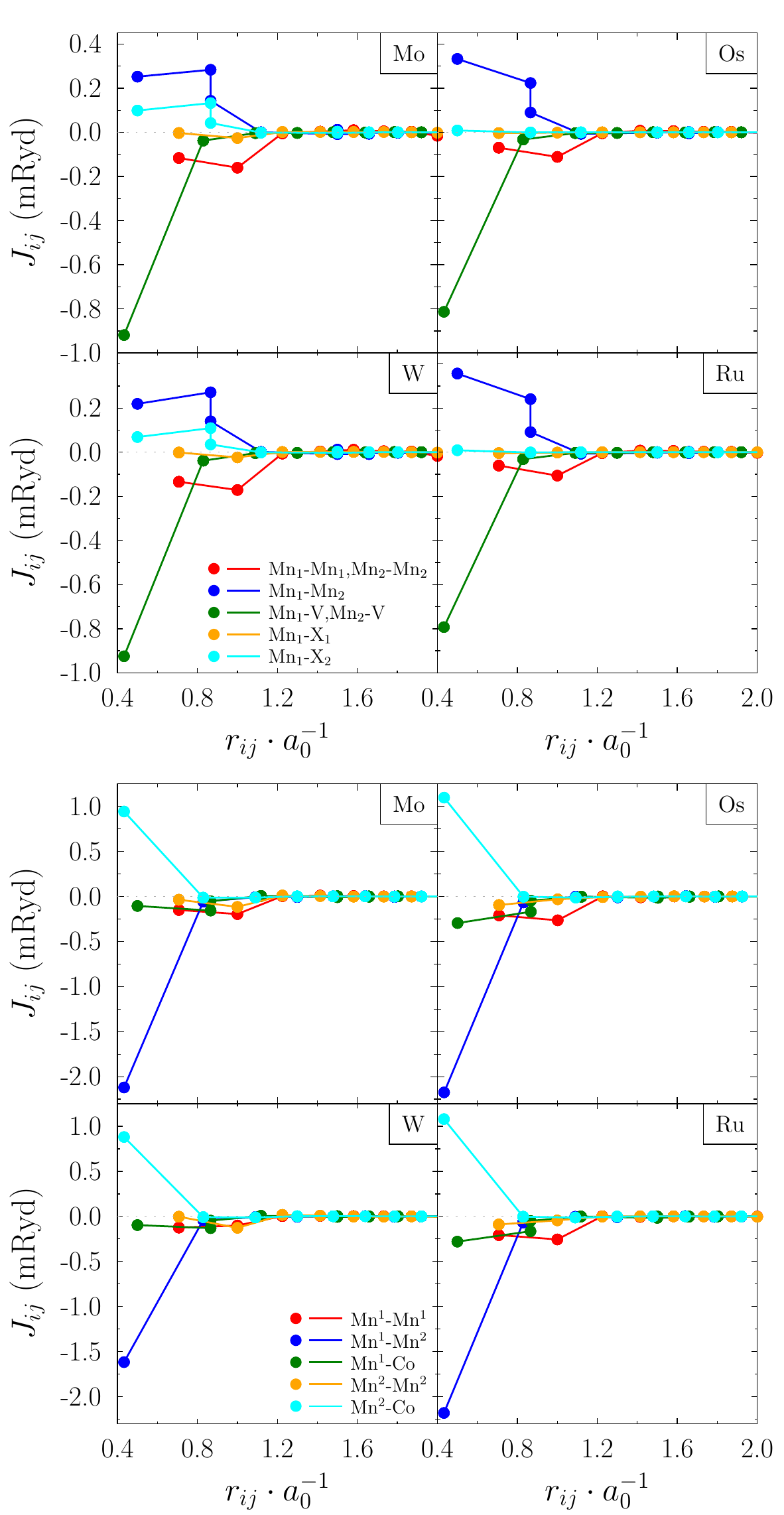}
\caption{(Color online) Intersublattice Heisenberg exchange parameter as a function of renormalized interatomic distance for a) Mn$_{2-x}$\textit{Z}$_{x}$VAl and b) Mn$_{2-x}$\textit{Z}$_{x}$CoAl, where the different subpanels show the dopants W (bottom left), Ru (bottom right), Mo (top left), and Os (top right). Different colours represents the coupling between Mn$^1$-Mn$^1$ (red dotes), Mn$^1$-Mn$^2$ (blue dotes), Mn$^1$-Co or Mn$^1$-V (green dotes), Mn$^2$-Mn$^2$ (orange dotes) and Mn$^2$-Co or Mn$^2$-V (cyan dotes).}
\label{fig:5}
\end{figure}

As shown in Fig.~\ref{fig:5}, doping with 4d and 5d elements reduces nearest-neighbour interactions and the correlation length between magnetic moments, which is a direct consequence of the disorder and the coherent potential approximation \cite{Bottcher:2012hz}. Nearest neighbour interactions are affected mostly by the doping. In general, the exchange couplings diminish with doping concentration $x$ up to $\unit[0.6]{mRyd}$ for W and $x=0.1$. For Os and Ru doping, there is a slight increase of the exchange coupling (about $\unit[0.03]{mRyd}$).

\begin{figure}[ht]
\centering
\includegraphics[width=0.8\columnwidth]{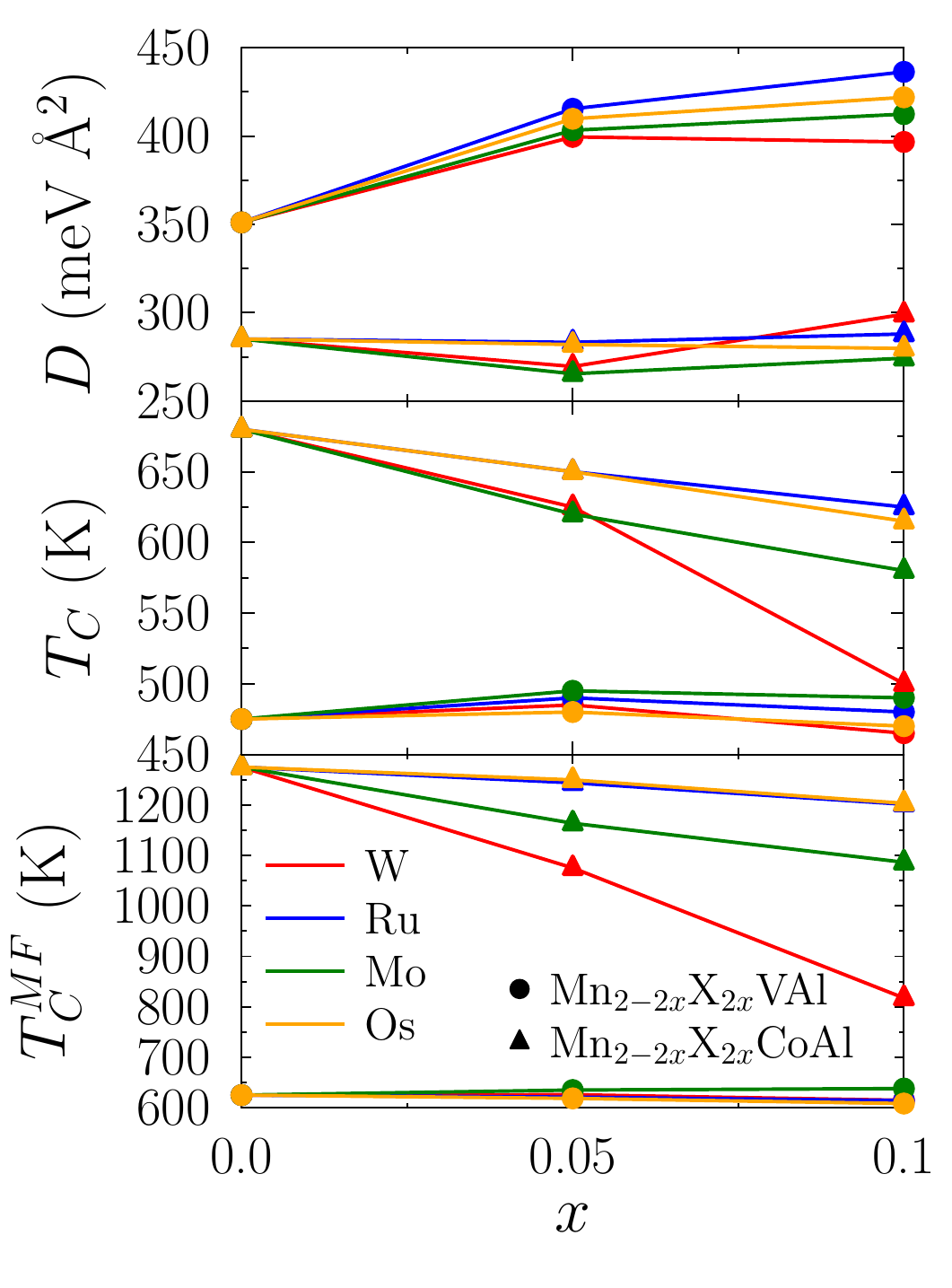}
\caption{(Color online) Spin wave stiffness $D$, critical temperatures $T_C$, and mean field critical temperatures $T_C^{MF}$ of Mn$_{2-x}$\textit{Z}$_{x}$CoAl (triangles) and Mn$_{2-x}$\textit{Z}$_{x}$VAl (circles) as a function of dopand concentration $x$. Dopands are Mn (black circles), Os (red squares), Ru (green diamonds), and W (orange triangles). }
\label{fig:6}
\end{figure}

With knowledge about the trends in the exchange couplings $\{J\}$, one can estimate the spin-wave stiffness $D$ and the phase transition temperature from both mean field theory via $k_B T_C^{MF}=\nicefrac{3}{2}\sum_j J_{0j}$ or from Monte Carlo simulations. The results are shown in Fig.~\ref{fig:6}. The spin-wave stiffness (upper panel in Fig.~\ref{fig:6}) for Mn$_{2-x}$CoAl is in good agreement with Ref.~[\onlinecite{Jakobsson:2015dt}], while for Mn$_{2-x}$VAl we reproduce the spin wave stiffness constant $D$ already reported in Ref.~[\onlinecite{Chico:2016dya}] ($D=\unit[324]{meV\AA^2}$), but not the experimentally measured stiffness \cite{Umetsu:2015db} ($D=\unit[534]{meV\AA^2}$ ). For the Co based Heusler compounds we obtain a hardening of the spin-waves after an initial softening, where for the V based Heusler compound, only hardening of the spin-waves with doping is observed. The phase transition temperature $T_C$, which turns out to be inversely proportional to $D$, decreases with doping concentration $x$ for two reasons, namely: \textit{i)} reduction of the magnetic moment due to doping and, consequently, stronger fluctuations at a given temperature as well as \textit{ii)} reduction of correlation. The critical temperature $T_C$ is obtained from Monte Carlo simulations on the Metropolis algorithm \cite{Anonymous:ypogcH9U}, from Binder's fourth cumulant \cite{Anonymous:ypogcH9U} for different simulated system sizes but also from the spin susceptibility $\chi$. Note that the first method could fail for antiferro- and ferrimagnets. Thus, we obtain a systematic error of about $\pm\unit[5]{K}$. 

Our simulations of ordering temperature ($\unit[680]{K}$ for Mn$_{2}$CoAl and $\unit[475]{K}$ for Mn$_{2}$VAl) underestimate the transition temperature observed from experiment ($\unit[720]{K}$ for Mn$_{2}$CoAl\cite{Ouardi:2013ko} and $\unit[768]{K}$ for Mn$_{2}$VAl). This discrepancy that is most notable for Mn$_{2-x}$VAl was reported earlier \cite{Chico:2016dya} and could have multiple reasons. First, magnetic properties in Heusler alloys are sensitive to the interstitial region spanned by the muffin tin potential. Thus, full-potential simulations are required as it was shown in Refs.~[\onlinecite{Liu:2008dt,Meinel:1995ct,Ouardi:2013ko}]. Also the results depend crucially on the choice of the exchange-correlation functional and on electron correlations e.g. addressed by including a Hubbard $U$\cite{Chico:2016dya}.  Second, the Heisenberg exchange is calculated for a collinear ferrimagnetic state but when the magnetic disorder is taken into account in the electronic structure, usually the exchange interaction is biased\cite{Bottcher:2012hz}.   Based on the alloy analogy model \cite{Ebert:2015kxa}, we modelled also the temperature stability of the magnetic properties (magnetic moments and magnetic exchange) coming from electronic structure by the partial disordered local moment (DLM) approximation within the Ising model \cite{Bottcher:2012hz}. DLM approach is believed to accurately describe `spin temperature' in the electronic structure \cite{Staunton:2013ww}. However, it turned out that the disordered local moment theory can not be applied to both Heusler and inverse Heusler for similar reasons as for Ni\cite{Akai:1993jy}: the magnetic moments in Al and Co/V disappear. For Mn$_2$CoAl, our simulations show furthermore that the magnetic moment of the Mn$^2$ atom is zero in the paramagnetic phase and, consequently, the magnetic exchange and the phase transition temperature are zero. This result is independent of the doping with 4d and 5d elements. These results indicate the inconsistency of the DLM model for Heusler materials. It is still an open question, if results get improved by applying relativistic DLM theory\cite{Buruzs:Pim91Guq}. Third, we consider only a simplified approach for electron correlation in the LDA and GGA density functional. However, it is known \cite{Chico:2016dya} that improved models for electron correlation have the trend to increase slightly the phase transition temperature.
	   
\subsection{Gilbert damping}

\begin{figure}[!h]
\centering
\includegraphics[width=1.0\columnwidth]{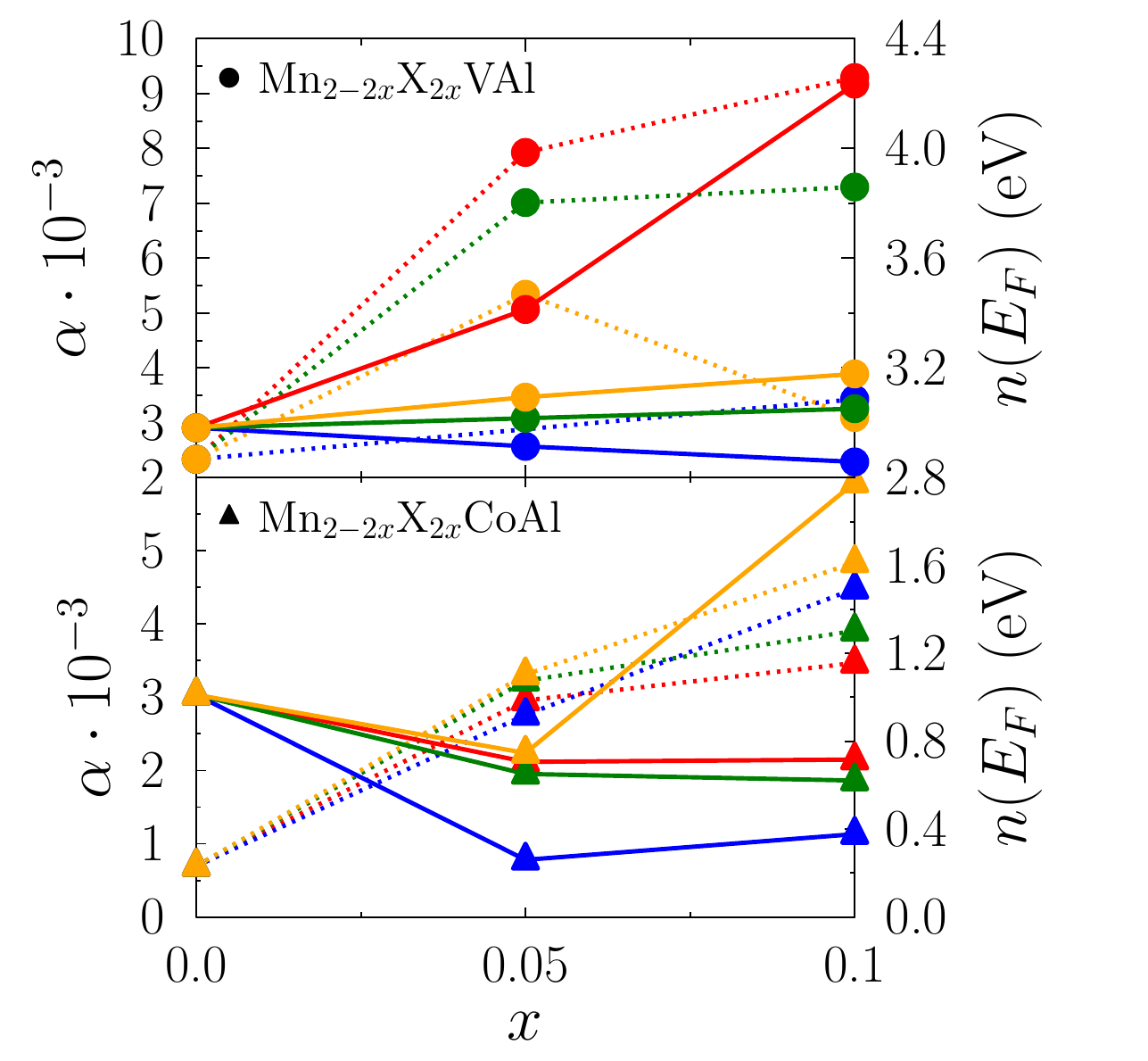}
\caption{(Color online) Gilbert damping parameters $\alpha$ (solid lines) and density of states at the Fermi level $n(E_F)$ (dotted lines) of Mn$_{2-x}$\textit{Z}$_{x}$CoAl (triangular symboles) and Mn$_{2-x}$\textit{Z}$_{x}$VAl (circle symboles) vs dopand concentration $x$. Dopands are W (red color), Ru (blue color), Mo (green color), and Os (orange color). }
\label{fig:7}
\end{figure}

Previous studies \cite{Chimata:2017iu} have shown that Gilbert damping is a crucial parameter in the ultrafast switching procedure and, thus, call for ab-initio footing. Figure~\ref{fig:7} shows the Gilbert damping $\alpha$ as a function $x$ at $T=\unit[300]{K}$. Note that for these calculations both lattice and magnetic fluctuations terms are considered, where the magnetic fluctuations are assumed from a linear correlation between the magnetization and the temperature. This could result in errors, in particular at high temperatures.

The Gilbert damping of both undoped Heusler materials (Mn$_{2}$CoAl: $\alpha=0.0030$, Mn$_{2}$VAl: $\alpha=0.0029$) is high compared to other half-metals reported, e.g., in Ref.~[\onlinecite{Chico:2016dya}] or low-damping alloys like Fe$_{0.75}$Co$_{0.25}$\cite{Schoen:2016gcc}. The trends of the Gilbert damping parameters with dopant concentration are different for Heusler and inverse Heusler materials. In Mn$_{2-x}$\textit{Z}$_{x}$VAl, doping leads to an increase of the damping with $x$, except for the case of Ru. The slope of $\alpha$ versus concentration $x$ follows the general increase of the total density of states at the Fermi level as it is proposed in Refs.~[\onlinecite{Ebert:2011gx,Lounis:2015ho,Schoen:2016gcc}], but not linear to it. This non-linearity was already observed for Heusler materials in Ref.~[\onlinecite{Chico:2016dya}] or doped permalloy with the heavy 4d and 5d elements used here \cite{Pan:2016gb}. The observed damping $\alpha$ is different from zero, however, small. This is in line with the theory proposed in Ref.~[\onlinecite{Lounis:2015ho}], in which damping is proportional to the product of the spin-polarised DOS and, consequently $\alpha\approx 0$. The increase of damping can be also understood in terms of the Kambersk\'y model \cite{Kambersky:1984iz,Kambersky:1976gi}: Alloying broadens the electron bands and more spin-flip transitions between the electron states occur. This is true only, if interband transitions are already dominating. In the inverse Heusler material Mn$_{2}$CoAl we even find a decrease with $x$. This is due to the spin-gapless semiconducting behaviour (cf. Fig.~\ref{fig:2}): Only a low number of states exist at the Fermi energy, making interband transitions unlikely. The damping is dominated by intraband transitions, that tend to decrease with very small $x$. With increasing $x$, however, states appear within the gap and interband transition are preferred. Thus, a small increase with even higher concentration is expected and observed. However, not only the number of states at the Fermi energy and the spectral width of the states contribute to the damping, but also the spin-orbit coupling (SOC), the Land\'e factor, and the saturation magnetization affect the damping parameter. Since we dope with rather heavy elements W, Mo, Ru, and Os, spin orbit coupling strongly contributes to the variation of damping with concentration $x$: the higher the `mass' of the dopant atom (W and Os compared to Ru and Mo) is, the higher is the damping parameter.

After we addressed all relevant parameters for the simulation based on the Landau-Lifshitz-Gilbert equation, we are able to perform ultrafast switching calculations.
   
\subsection{Ultrafast switching}
\label{sec:dynamics}

\begin{figure}[!ht]
\centering
\includegraphics[width=0.7\columnwidth]{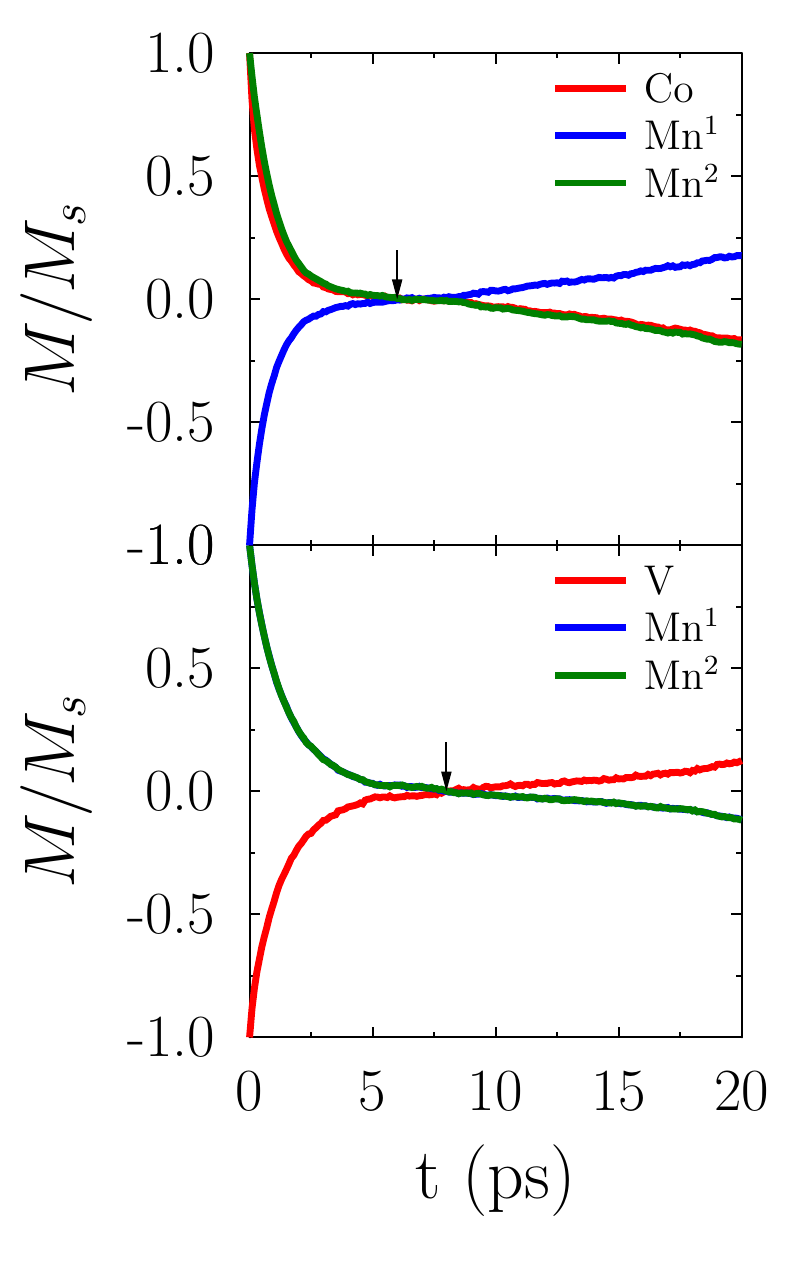}
\caption{(Color online) Ultrafast switching behaviour of Mn$_{2}$CoAl (upper panel) and Mn$_{2}$VAl (lower panel). The demagnetization is shown element resolved (blue and green lines - Mn atoms, red line - Co/ V atom). The peak temperature is $\unit[600]{K}$ for Mn$_{2}$VAl and $\unit[900]{K}$ for Mn$_{2}$CoAl. The external magnetic field is $B=\unit[2.5]{T}$ and  damping parameter is $\alpha=0.009$. The arrow indicates the crossing point at where the switching takes place.} 
\label{fig:8}
\end{figure}

\begin{figure}
\centering
\includegraphics[width=0.9\columnwidth]{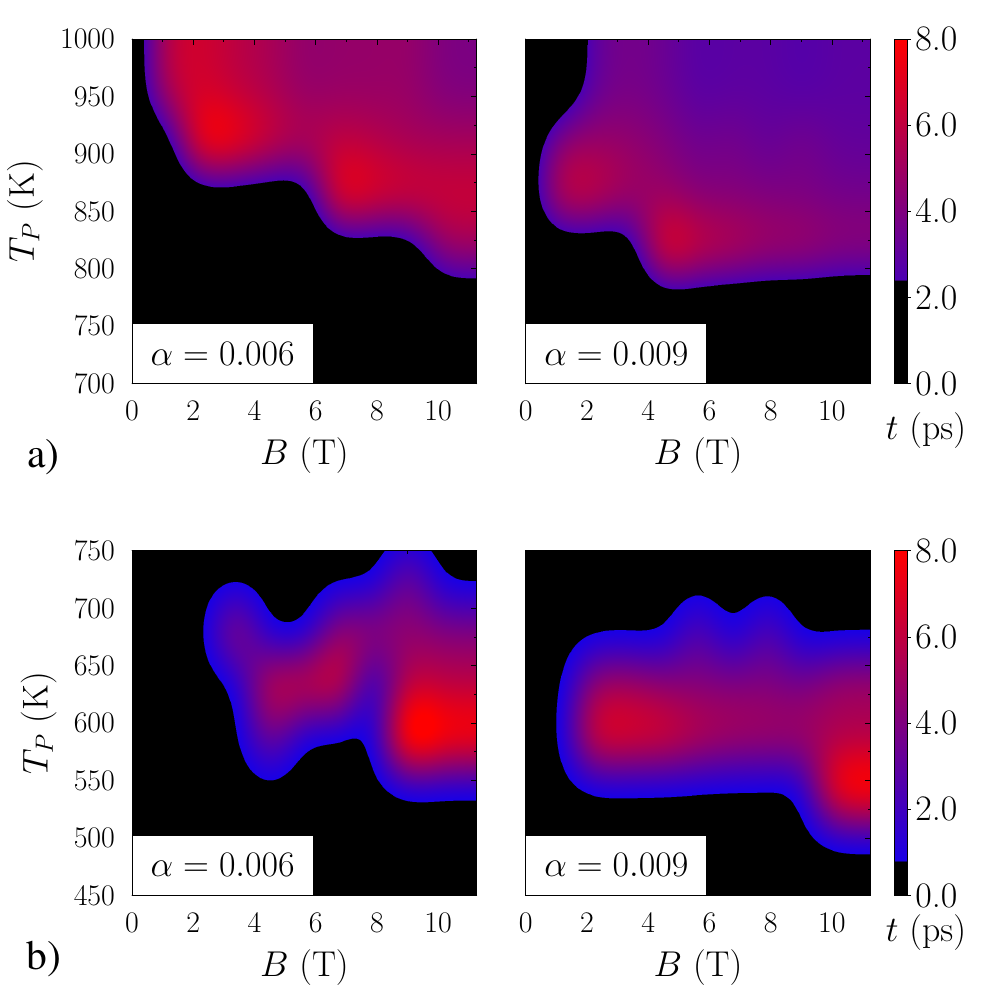}
\caption{(Color online) Thermal switching phase diagram for different damping parameter (0.006 and 0.009) in a) Mn$_{2}$CoAl and b) Mn$_2$VAl.  The peak temperature is represented versus the strength of the external magnetic field. The colour scale (fast switching - blue colour, slow switching - red colour) represents the time in units of ps where the switching (indicated by an arrow in Fig.~(\ref{fig:8})) takes place. No switching is represented by the black background. }
\label{fig:9}
\end{figure}

In order to study the ultrafast switching process in Heusler alloys we combined the two temperature model with an atomistic spin dynamics code~\cite{Skubic:2008gs}. Here, we considered a very long thermal pulse of $\unit[20]{ps}$ with different peak temperatures $T_P$. Typical timescales of the ultrafast demagnetization and remagnetization process for Mn$_2$VAl and Mn$_2$CoAl are in the orders of picoseconds ($\unit[1-5]{ps}$) (see Fig.~\ref{fig:8}). The time scales are mainly dictated by the Gilbert damping $\alpha$, which is varied in our studies between $0.003$, $0.006$, and $0.009$, but can depend on the Heisenberg exchange \cite{Mentink:2012ga}. As demonstrated above, these damping values are achievable by doping the `pure' Heusler materials. There is only a slight shift observable in the demagnetization time of each individual element in Mn$_2$CoAl, where for Mn$_2$VAl, it is not. After demagnetization, the Heusler material undergoes reliable switching only when an external magnetic field induced by the pump-pulse is present. Thus, three parameters --- damping, peak temperature and pulse induced external magnetic field --- span a phase space for observing reliable switching, as shown in Fig.~\ref{fig:9}. 

We did not observe any magnetic switching for both Heusler materials with $\alpha=0.003$ (data not shown here). Typically for certain threshold peak temperatures $T_P$ above the magnetic phase transition temperature ($T_C=\unit[700]{K}$ for Mn$_2$CoAl and $T_C=\unit[475]{K}$ for Mn$_2$VAl) switching occurs. The peak temperature can be tuned by the laser intensity and the pulse duration. The presence of an effective magnetic field during pumping is discussed in literature \cite{Berritta:2016em,2017NatSR...7.4114J}. It was argued that the electric field of the pump pulse induces a strong material specific magnetic field of $\unit[10-100]{T}$. Even below but above certain minimum magnetic field of $\unit[1-2]{T}$, we observed reliable switching. This threshold magnetic field as well as the switching time (indicated by reduced contrast in Fig.~\ref{fig:9}) decreases with increasing damping. The time when the switching occurs (crossing point in Fig.~\ref{fig:8} and colour scale in Fig.~\ref{fig:9}) typically passes a maximum at certain and decreases for larger peak temperatures. However, there is also a minimum switching time of around $\unit[2-3]{ps}$, controlled by the demagnetization rate. Note that due to the different spin polarization and resulting different atomic magnetic moments and magnetic states, an asymmetry in the phase diagram between Mn$_2$CoAl and Mn$_2$VAl occurs.
 
Nevertheless, our approach has certain limitations. For instance, we explicitly neglect the electronic motion and effects like super diffusion or spin-flip scattering, as discussed in Ref. \cite{Battiato:2010br}. We also assume the damping to be `spin- and phonon-temperature' independent. This is a rough approximation, in particular, due to the important role of phonons in the demagnetization process (e.g. Ref.~[\onlinecite{Illg:2013jp}]) and for energy dissipation in magnetic systems \cite{Ebert:2011gx}. Furthermore, we neglect the change of the magnetic exchange interaction with temperature, although magnetic moments of Co and V atoms vanish in the DLM approximation. This behaviour in the disordered local moment theory is well studied \cite{Akai:1993jy} and occurs also for Ni atoms. But we have shown elsewhere\cite{Chimata:2012kv} but also others \cite{Evans:2015ffa,Hinzke:2015kra,Atxitia:2010fu,Hinzke:2015kra}, that our methodology is applicable for demagnetization in bulk bcc Fe and hcp Co compounds and, likely, for the Heusler materials studied here. We also neglect possible structural phase transition to A$2$ or $B2$ disorder during demagnetization.

\section{Conclusion}
\label{conclusion}
We have demonstrated thermal switching in Heusler and inverse Heusler materials making use of magnetic field pulse induced by the pump-pulse. We found a sensitive dependence of the possible switching and the switching time on the magnetic field pulse strength, the peak temperature in the effective two-temperature model as well as intrinsic materials properties, say the Heisenberg exchange and the Gilbert damping parameter. We have shown that the latter can be tuned by doping heavy elements, say W, Mo, Ru, Os, to both, higher and lower damping values, especially in the case of spin-gapless semiconductor. This calls for further investigations on other spin-gapless semiconductor~\cite{Jakobsson:2015dt}, aiming for tuning the Gilbert damping to very low values, which may enable interesting spintronic and magnonic applications~\cite{Palmstrom:2016ju}. Within our methodology, we could reproduce exchange parameter and, consequently, phase transition temperatures reported in literature~\cite{Jakobsson:2015dt}. Our overall finding extends with Heusler and inverse Heusler alloys the class of materials that exhibits laser induced magnetic switching and calls for future theoretical and experimental studies.

\section{Acknowledgement}
We acknowledge financial support from the Swedish Research Council. O.E. and and E.K.D.-Cz. acknowledged support from KAW (projects 2013.0020 and 2012.0031) as well as acknowledges eSSENCE and STandUP. The calculations were performed at NSC (Link\"oping University, Sweden) under a SNAC project.

\bibliographystyle{apsrev}

\begin{thebibliography}{85}
\expandafter\ifx\csname natexlab\endcsname\relax\def\natexlab#1{#1}\fi
\expandafter\ifx\csname bibnamefont\endcsname\relax
  \def\bibnamefont#1{#1}\fi
\expandafter\ifx\csname bibfnamefont\endcsname\relax
  \def\bibfnamefont#1{#1}\fi
\expandafter\ifx\csname citenamefont\endcsname\relax
  \def\citenamefont#1{#1}\fi
\expandafter\ifx\csname url\endcsname\relax
  \def\url#1{\texttt{#1}}\fi
\expandafter\ifx\csname urlprefix\endcsname\relax\def\urlprefix{URL }\fi
\providecommand{\bibinfo}[2]{#2}
\providecommand{\eprint}[2][]{\url{#2}}

\bibitem[{\citenamefont{Beaurepaire et~al.}(1996)\citenamefont{Beaurepaire,
  Merle, Daunois, and Bigot}}]{Beaurepaire:1996es}
\bibinfo{author}{\bibfnamefont{E.}~\bibnamefont{Beaurepaire}},
  \bibinfo{author}{\bibfnamefont{J.~C.} \bibnamefont{Merle}},
  \bibinfo{author}{\bibfnamefont{A.}~\bibnamefont{Daunois}}, \bibnamefont{and}
  \bibinfo{author}{\bibfnamefont{J.~Y.} \bibnamefont{Bigot}},
  \bibinfo{journal}{Phys. Rev. Lett.} \textbf{\bibinfo{volume}{76}},
  \bibinfo{pages}{4250} (\bibinfo{year}{1996}),
  \urlprefix\url{https://link.aps.org/doi/10.1103/PhysRevLett.76.4250}.

\bibitem[{\citenamefont{Carpene et~al.}(2008)\citenamefont{Carpene, Mancini,
  Dallera, Brenna, Puppin, and De~Silvestri}}]{Carpene:2008bd}
\bibinfo{author}{\bibfnamefont{E.}~\bibnamefont{Carpene}},
  \bibinfo{author}{\bibfnamefont{E.}~\bibnamefont{Mancini}},
  \bibinfo{author}{\bibfnamefont{C.}~\bibnamefont{Dallera}},
  \bibinfo{author}{\bibfnamefont{M.}~\bibnamefont{Brenna}},
  \bibinfo{author}{\bibfnamefont{E.}~\bibnamefont{Puppin}}, \bibnamefont{and}
  \bibinfo{author}{\bibfnamefont{S.}~\bibnamefont{De~Silvestri}},
  \bibinfo{journal}{Phys. Rev. B} \textbf{\bibinfo{volume}{78}},
  \bibinfo{pages}{174422} (\bibinfo{year}{2008}),
  \urlprefix\url{https://link.aps.org/doi/10.1103/PhysRevB.78.174422}.

\bibitem[{\citenamefont{Cinchetti et~al.}(2006)\citenamefont{Cinchetti,
  Albaneda, Hoffmann, Roth, W{\"u}stenberg, Krau{\ss}, Andreyev, Schneider,
  Bauer, and Aeschlimann}}]{Cinchetti:2006cv}
\bibinfo{author}{\bibfnamefont{M.}~\bibnamefont{Cinchetti}},
  \bibinfo{author}{\bibfnamefont{M.~S.} \bibnamefont{Albaneda}},
  \bibinfo{author}{\bibfnamefont{D.}~\bibnamefont{Hoffmann}},
  \bibinfo{author}{\bibfnamefont{T.}~\bibnamefont{Roth}},
  \bibinfo{author}{\bibfnamefont{J.~P.} \bibnamefont{W{\"u}stenberg}},
  \bibinfo{author}{\bibfnamefont{M.}~\bibnamefont{Krau{\ss}}},
  \bibinfo{author}{\bibfnamefont{O.}~\bibnamefont{Andreyev}},
  \bibinfo{author}{\bibfnamefont{H.~C.} \bibnamefont{Schneider}},
  \bibinfo{author}{\bibfnamefont{M.}~\bibnamefont{Bauer}}, \bibnamefont{and}
  \bibinfo{author}{\bibfnamefont{M.}~\bibnamefont{Aeschlimann}},
  \bibinfo{journal}{Phys. Rev. Lett.} \textbf{\bibinfo{volume}{97}},
  \bibinfo{pages}{177201} (\bibinfo{year}{2006}),
  \urlprefix\url{https://link.aps.org/doi/10.1103/PhysRevLett.97.177201}.

\bibitem[{\citenamefont{Stamm et~al.}(2007)\citenamefont{Stamm, Kachel,
  Pontius, Mitzner, Quast, Holldack, Khan, Lupulescu, Aziz, Wietstruk
  et~al.}}]{2007NatMa...6..740S}
\bibinfo{author}{\bibfnamefont{C.}~\bibnamefont{Stamm}},
  \bibinfo{author}{\bibfnamefont{T.}~\bibnamefont{Kachel}},
  \bibinfo{author}{\bibfnamefont{N.}~\bibnamefont{Pontius}},
  \bibinfo{author}{\bibfnamefont{R.}~\bibnamefont{Mitzner}},
  \bibinfo{author}{\bibfnamefont{T.}~\bibnamefont{Quast}},
  \bibinfo{author}{\bibfnamefont{K.}~\bibnamefont{Holldack}},
  \bibinfo{author}{\bibfnamefont{S.}~\bibnamefont{Khan}},
  \bibinfo{author}{\bibfnamefont{C.}~\bibnamefont{Lupulescu}},
  \bibinfo{author}{\bibfnamefont{E.~F.} \bibnamefont{Aziz}},
  \bibinfo{author}{\bibfnamefont{M.}~\bibnamefont{Wietstruk}},
  \bibnamefont{et~al.}, \bibinfo{journal}{Nature Materials}
  \textbf{\bibinfo{volume}{6}}, \bibinfo{pages}{740} (\bibinfo{year}{2007}),
  \urlprefix\url{http://adsabs.harvard.edu/cgi-bin/nph-data_query?bibcode=2007NatMa...6..740S&link_type=EJOURNAL}.

\bibitem[{\citenamefont{Rhie et~al.}(2003)\citenamefont{Rhie, D{\"u}rr, and
  Eberhardt}}]{Rhie:2003ji}
\bibinfo{author}{\bibfnamefont{H.~S.} \bibnamefont{Rhie}},
  \bibinfo{author}{\bibfnamefont{H.~A.} \bibnamefont{D{\"u}rr}},
  \bibnamefont{and}
  \bibinfo{author}{\bibfnamefont{W.}~\bibnamefont{Eberhardt}},
  \bibinfo{journal}{Phys. Rev. Lett.} \textbf{\bibinfo{volume}{90}},
  \bibinfo{pages}{247201} (\bibinfo{year}{2003}),
  \urlprefix\url{https://link.aps.org/doi/10.1103/PhysRevLett.90.247201}.

\bibitem[{\citenamefont{Wietstruk et~al.}(2011)\citenamefont{Wietstruk,
  Melnikov, Stamm, Kachel, Pontius, Sultan, Gahl, Weinelt, D{\"u}rr, and
  Bovensiepen}}]{Wietstruk:2011ex}
\bibinfo{author}{\bibfnamefont{M.}~\bibnamefont{Wietstruk}},
  \bibinfo{author}{\bibfnamefont{A.}~\bibnamefont{Melnikov}},
  \bibinfo{author}{\bibfnamefont{C.}~\bibnamefont{Stamm}},
  \bibinfo{author}{\bibfnamefont{T.}~\bibnamefont{Kachel}},
  \bibinfo{author}{\bibfnamefont{N.}~\bibnamefont{Pontius}},
  \bibinfo{author}{\bibfnamefont{M.}~\bibnamefont{Sultan}},
  \bibinfo{author}{\bibfnamefont{C.}~\bibnamefont{Gahl}},
  \bibinfo{author}{\bibfnamefont{M.}~\bibnamefont{Weinelt}},
  \bibinfo{author}{\bibfnamefont{H.~A.} \bibnamefont{D{\"u}rr}},
  \bibnamefont{and}
  \bibinfo{author}{\bibfnamefont{U.}~\bibnamefont{Bovensiepen}},
  \bibinfo{journal}{Phys. Rev. Lett.} \textbf{\bibinfo{volume}{106}},
  \bibinfo{pages}{127401} (\bibinfo{year}{2011}),
  \urlprefix\url{https://link.aps.org/doi/10.1103/PhysRevLett.106.127401}.

\bibitem[{\citenamefont{Stanciu et~al.}(2006)\citenamefont{Stanciu, Kimel,
  Hansteen, Tsukamoto, Itoh, Kirilyuk, and Rasing}}]{Stanciu:2006gm}
\bibinfo{author}{\bibfnamefont{C.~D.} \bibnamefont{Stanciu}},
  \bibinfo{author}{\bibfnamefont{A.~V.} \bibnamefont{Kimel}},
  \bibinfo{author}{\bibfnamefont{F.}~\bibnamefont{Hansteen}},
  \bibinfo{author}{\bibfnamefont{A.}~\bibnamefont{Tsukamoto}},
  \bibinfo{author}{\bibfnamefont{A.}~\bibnamefont{Itoh}},
  \bibinfo{author}{\bibfnamefont{A.}~\bibnamefont{Kirilyuk}}, \bibnamefont{and}
  \bibinfo{author}{\bibfnamefont{T.}~\bibnamefont{Rasing}},
  \bibinfo{journal}{Phys. Rev. B} \textbf{\bibinfo{volume}{73}},
  \bibinfo{pages}{220402} (\bibinfo{year}{2006}),
  \urlprefix\url{https://link.aps.org/doi/10.1103/PhysRevB.73.220402}.

\bibitem[{\citenamefont{Vahaplar et~al.}(2009)\citenamefont{Vahaplar,
  Kalashnikova, Kimel, Hinzke, Nowak, Chantrell, Tsukamoto, Itoh, Kirilyuk, and
  Rasing}}]{Vahaplar:2009fpa}
\bibinfo{author}{\bibfnamefont{K.}~\bibnamefont{Vahaplar}},
  \bibinfo{author}{\bibfnamefont{A.~M.} \bibnamefont{Kalashnikova}},
  \bibinfo{author}{\bibfnamefont{A.~V.} \bibnamefont{Kimel}},
  \bibinfo{author}{\bibfnamefont{D.}~\bibnamefont{Hinzke}},
  \bibinfo{author}{\bibfnamefont{U.}~\bibnamefont{Nowak}},
  \bibinfo{author}{\bibfnamefont{R.}~\bibnamefont{Chantrell}},
  \bibinfo{author}{\bibfnamefont{A.}~\bibnamefont{Tsukamoto}},
  \bibinfo{author}{\bibfnamefont{A.}~\bibnamefont{Itoh}},
  \bibinfo{author}{\bibfnamefont{A.}~\bibnamefont{Kirilyuk}}, \bibnamefont{and}
  \bibinfo{author}{\bibfnamefont{T.}~\bibnamefont{Rasing}},
  \bibinfo{journal}{Phys. Rev. Lett.} \textbf{\bibinfo{volume}{103}},
  \bibinfo{pages}{117201} (\bibinfo{year}{2009}),
  \urlprefix\url{https://link.aps.org/doi/10.1103/PhysRevLett.103.117201}.

\bibitem[{\citenamefont{Stanciu et~al.}(2007)\citenamefont{Stanciu, Hansteen,
  Kimel, Kirilyuk, Tsukamoto, Itoh, and Rasing}}]{Stanciu:2007fy}
\bibinfo{author}{\bibfnamefont{C.~D.} \bibnamefont{Stanciu}},
  \bibinfo{author}{\bibfnamefont{F.}~\bibnamefont{Hansteen}},
  \bibinfo{author}{\bibfnamefont{A.~V.} \bibnamefont{Kimel}},
  \bibinfo{author}{\bibfnamefont{A.}~\bibnamefont{Kirilyuk}},
  \bibinfo{author}{\bibfnamefont{A.}~\bibnamefont{Tsukamoto}},
  \bibinfo{author}{\bibfnamefont{A.}~\bibnamefont{Itoh}}, \bibnamefont{and}
  \bibinfo{author}{\bibfnamefont{T.}~\bibnamefont{Rasing}},
  \bibinfo{journal}{Phys. Rev. Lett.} \textbf{\bibinfo{volume}{99}},
  \bibinfo{pages}{047601} (\bibinfo{year}{2007}),
  \urlprefix\url{https://link.aps.org/doi/10.1103/PhysRevLett.99.047601}.

\bibitem[{\citenamefont{Steil et~al.}(2011)\citenamefont{Steil, Alebrand,
  Hassdenteufel, Cinchetti, and Aeschlimann}}]{Steil:2011dt}
\bibinfo{author}{\bibfnamefont{D.}~\bibnamefont{Steil}},
  \bibinfo{author}{\bibfnamefont{S.}~\bibnamefont{Alebrand}},
  \bibinfo{author}{\bibfnamefont{A.}~\bibnamefont{Hassdenteufel}},
  \bibinfo{author}{\bibfnamefont{M.}~\bibnamefont{Cinchetti}},
  \bibnamefont{and}
  \bibinfo{author}{\bibfnamefont{M.}~\bibnamefont{Aeschlimann}},
  \bibinfo{journal}{Phys. Rev. B} \textbf{\bibinfo{volume}{84}},
  \bibinfo{pages}{224408} (\bibinfo{year}{2011}),
  \urlprefix\url{https://link.aps.org/doi/10.1103/PhysRevB.84.224408}.

\bibitem[{\citenamefont{Ostler et~al.}(2012)\citenamefont{Ostler, Barker,
  Evans, Chantrell, Atxitia, Chubykalo-Fesenko, El~Moussaoui, Le~Guyader,
  Mengotti, Heyderman et~al.}}]{Ostler:2012hx}
\bibinfo{author}{\bibfnamefont{T.~A.} \bibnamefont{Ostler}},
  \bibinfo{author}{\bibfnamefont{J.}~\bibnamefont{Barker}},
  \bibinfo{author}{\bibfnamefont{R.~F.~L.} \bibnamefont{Evans}},
  \bibinfo{author}{\bibfnamefont{R.~W.} \bibnamefont{Chantrell}},
  \bibinfo{author}{\bibfnamefont{U.}~\bibnamefont{Atxitia}},
  \bibinfo{author}{\bibfnamefont{O.}~\bibnamefont{Chubykalo-Fesenko}},
  \bibinfo{author}{\bibfnamefont{S.}~\bibnamefont{El~Moussaoui}},
  \bibinfo{author}{\bibfnamefont{L.}~\bibnamefont{Le~Guyader}},
  \bibinfo{author}{\bibfnamefont{E.}~\bibnamefont{Mengotti}},
  \bibinfo{author}{\bibfnamefont{L.~J.} \bibnamefont{Heyderman}},
  \bibnamefont{et~al.}, \bibinfo{journal}{Nat Comms}
  \textbf{\bibinfo{volume}{3}}, \bibinfo{pages}{666} (\bibinfo{year}{2012}),
  \urlprefix\url{https://www.nature.com/articles/ncomms1666}.

\bibitem[{\citenamefont{Mentink et~al.}(2012)\citenamefont{Mentink, Hellsvik,
  Afanasiev, Ivanov, Kirilyuk, Kimel, Eriksson, Katsnelson, and
  Rasing}}]{Mentink:2012ga}
\bibinfo{author}{\bibfnamefont{J.~H.} \bibnamefont{Mentink}},
  \bibinfo{author}{\bibfnamefont{J.}~\bibnamefont{Hellsvik}},
  \bibinfo{author}{\bibfnamefont{D.~V.} \bibnamefont{Afanasiev}},
  \bibinfo{author}{\bibfnamefont{B.~A.} \bibnamefont{Ivanov}},
  \bibinfo{author}{\bibfnamefont{A.}~\bibnamefont{Kirilyuk}},
  \bibinfo{author}{\bibfnamefont{A.~V.} \bibnamefont{Kimel}},
  \bibinfo{author}{\bibfnamefont{O.}~\bibnamefont{Eriksson}},
  \bibinfo{author}{\bibfnamefont{M.~I.} \bibnamefont{Katsnelson}},
  \bibnamefont{and} \bibinfo{author}{\bibfnamefont{T.}~\bibnamefont{Rasing}},
  \bibinfo{journal}{Phys. Rev. Lett.} \textbf{\bibinfo{volume}{108}},
  \bibinfo{pages}{057202} (\bibinfo{year}{2012}),
  \urlprefix\url{https://link.aps.org/doi/10.1103/PhysRevLett.108.057202}.

\bibitem[{\citenamefont{Chimata et~al.}(2015)\citenamefont{Chimata, Isaeva,
  K{\'a}das, Bergman, Sanyal, Mentink, Katsnelson, Rasing, Kirilyuk, Kimel
  et~al.}}]{Chimata:2015bea}
\bibinfo{author}{\bibfnamefont{R.}~\bibnamefont{Chimata}},
  \bibinfo{author}{\bibfnamefont{L.}~\bibnamefont{Isaeva}},
  \bibinfo{author}{\bibfnamefont{K.}~\bibnamefont{K{\'a}das}},
  \bibinfo{author}{\bibfnamefont{A.}~\bibnamefont{Bergman}},
  \bibinfo{author}{\bibfnamefont{B.}~\bibnamefont{Sanyal}},
  \bibinfo{author}{\bibfnamefont{J.~H.} \bibnamefont{Mentink}},
  \bibinfo{author}{\bibfnamefont{M.~I.} \bibnamefont{Katsnelson}},
  \bibinfo{author}{\bibfnamefont{T.}~\bibnamefont{Rasing}},
  \bibinfo{author}{\bibfnamefont{A.}~\bibnamefont{Kirilyuk}},
  \bibinfo{author}{\bibfnamefont{A.}~\bibnamefont{Kimel}},
  \bibnamefont{et~al.}, \bibinfo{journal}{Phys. Rev. B}
  \textbf{\bibinfo{volume}{92}}, \bibinfo{pages}{094411}
  (\bibinfo{year}{2015}),
  \urlprefix\url{https://link.aps.org/doi/10.1103/PhysRevB.92.094411}.

\bibitem[{\citenamefont{Alebrand et~al.}(2014)\citenamefont{Alebrand,
  Bierbrauer, Hehn, Gottwald, Schmitt, Steil, Fullerton, Mangin, Cinchetti, and
  Aeschlimann}}]{Alebrand:2014fp}
\bibinfo{author}{\bibfnamefont{S.}~\bibnamefont{Alebrand}},
  \bibinfo{author}{\bibfnamefont{U.}~\bibnamefont{Bierbrauer}},
  \bibinfo{author}{\bibfnamefont{M.}~\bibnamefont{Hehn}},
  \bibinfo{author}{\bibfnamefont{M.}~\bibnamefont{Gottwald}},
  \bibinfo{author}{\bibfnamefont{O.}~\bibnamefont{Schmitt}},
  \bibinfo{author}{\bibfnamefont{D.}~\bibnamefont{Steil}},
  \bibinfo{author}{\bibfnamefont{E.~E.} \bibnamefont{Fullerton}},
  \bibinfo{author}{\bibfnamefont{S.}~\bibnamefont{Mangin}},
  \bibinfo{author}{\bibfnamefont{M.}~\bibnamefont{Cinchetti}},
  \bibnamefont{and}
  \bibinfo{author}{\bibfnamefont{M.}~\bibnamefont{Aeschlimann}},
  \bibinfo{journal}{Phys. Rev. B} \textbf{\bibinfo{volume}{89}},
  \bibinfo{pages}{144404} (\bibinfo{year}{2014}),
  \urlprefix\url{https://link.aps.org/doi/10.1103/PhysRevB.89.144404}.

\bibitem[{\citenamefont{Bigot et~al.}(2009)\citenamefont{Bigot, Vomir, and
  Beaurepaire}}]{Bigot:2009bu}
\bibinfo{author}{\bibfnamefont{J.-Y.} \bibnamefont{Bigot}},
  \bibinfo{author}{\bibfnamefont{M.}~\bibnamefont{Vomir}}, \bibnamefont{and}
  \bibinfo{author}{\bibfnamefont{E.}~\bibnamefont{Beaurepaire}},
  \bibinfo{journal}{Nat Phys} \textbf{\bibinfo{volume}{5}},
  \bibinfo{pages}{515} (\bibinfo{year}{2009}),
  \urlprefix\url{https://www.nature.com/articles/nphys1285}.

\bibitem[{\citenamefont{M{\"u}ller et~al.}(2008)\citenamefont{M{\"u}ller,
  Walowski, Djordjevic, Miao, Gupta, Ramos, Gehrke, Moshnyaga, Samwer,
  Schmalhorst et~al.}}]{Muller:2008ip}
\bibinfo{author}{\bibfnamefont{G.~M.} \bibnamefont{M{\"u}ller}},
  \bibinfo{author}{\bibfnamefont{J.}~\bibnamefont{Walowski}},
  \bibinfo{author}{\bibfnamefont{M.}~\bibnamefont{Djordjevic}},
  \bibinfo{author}{\bibfnamefont{G.-X.} \bibnamefont{Miao}},
  \bibinfo{author}{\bibfnamefont{A.}~\bibnamefont{Gupta}},
  \bibinfo{author}{\bibfnamefont{A.~V.} \bibnamefont{Ramos}},
  \bibinfo{author}{\bibfnamefont{K.}~\bibnamefont{Gehrke}},
  \bibinfo{author}{\bibfnamefont{V.}~\bibnamefont{Moshnyaga}},
  \bibinfo{author}{\bibfnamefont{K.}~\bibnamefont{Samwer}},
  \bibinfo{author}{\bibfnamefont{J.}~\bibnamefont{Schmalhorst}},
  \bibnamefont{et~al.}, \bibinfo{journal}{Nature Publishing Group}
  \textbf{\bibinfo{volume}{8}}, \bibinfo{pages}{56} (\bibinfo{year}{2008}),
  \urlprefix\url{https://www.nature.com/articles/nmat2341}.

\bibitem[{\citenamefont{W{\"u}stenberg
  et~al.}(2011)\citenamefont{W{\"u}stenberg, Steil, Alebrand, Roth,
  Aeschlimann, and Cinchetti}}]{Wustenberg:2011co}
\bibinfo{author}{\bibfnamefont{J.~P.} \bibnamefont{W{\"u}stenberg}},
  \bibinfo{author}{\bibfnamefont{D.}~\bibnamefont{Steil}},
  \bibinfo{author}{\bibfnamefont{S.}~\bibnamefont{Alebrand}},
  \bibinfo{author}{\bibfnamefont{T.}~\bibnamefont{Roth}},
  \bibinfo{author}{\bibfnamefont{M.}~\bibnamefont{Aeschlimann}},
  \bibnamefont{and}
  \bibinfo{author}{\bibfnamefont{M.}~\bibnamefont{Cinchetti}},
  \bibinfo{journal}{Phys. Status Solidi B} \textbf{\bibinfo{volume}{248}},
  \bibinfo{pages}{2330} (\bibinfo{year}{2011}),
  \urlprefix\url{http://onlinelibrary.wiley.com/doi/10.1002/pssb.201147087/full}.

\bibitem[{\citenamefont{Mann et~al.}(2012)\citenamefont{Mann, Walowski,
  M{\"u}nzenberg, Maat, Carey, Childress, Mewes, Ebke, Drewello, Reiss
  et~al.}}]{Mann:2012bra}
\bibinfo{author}{\bibfnamefont{A.}~\bibnamefont{Mann}},
  \bibinfo{author}{\bibfnamefont{J.}~\bibnamefont{Walowski}},
  \bibinfo{author}{\bibfnamefont{M.}~\bibnamefont{M{\"u}nzenberg}},
  \bibinfo{author}{\bibfnamefont{S.}~\bibnamefont{Maat}},
  \bibinfo{author}{\bibfnamefont{M.~J.} \bibnamefont{Carey}},
  \bibinfo{author}{\bibfnamefont{J.~R.} \bibnamefont{Childress}},
  \bibinfo{author}{\bibfnamefont{C.}~\bibnamefont{Mewes}},
  \bibinfo{author}{\bibfnamefont{D.}~\bibnamefont{Ebke}},
  \bibinfo{author}{\bibfnamefont{V.}~\bibnamefont{Drewello}},
  \bibinfo{author}{\bibfnamefont{G.}~\bibnamefont{Reiss}},
  \bibnamefont{et~al.}, \bibinfo{journal}{Phys. Rev. X}
  \textbf{\bibinfo{volume}{2}}, \bibinfo{pages}{041008} (\bibinfo{year}{2012}),
  \urlprefix\url{https://link.aps.org/doi/10.1103/PhysRevX.2.041008}.

\bibitem[{\citenamefont{Steil et~al.}(2010)\citenamefont{Steil, Alebrand, Roth,
  Krau{\ss}, Kubota, Oogane, Ando, Schneider, Aeschlimann, and
  Cinchetti}}]{Steil:2010fi}
\bibinfo{author}{\bibfnamefont{D.}~\bibnamefont{Steil}},
  \bibinfo{author}{\bibfnamefont{S.}~\bibnamefont{Alebrand}},
  \bibinfo{author}{\bibfnamefont{T.}~\bibnamefont{Roth}},
  \bibinfo{author}{\bibfnamefont{M.}~\bibnamefont{Krau{\ss}}},
  \bibinfo{author}{\bibfnamefont{T.}~\bibnamefont{Kubota}},
  \bibinfo{author}{\bibfnamefont{M.}~\bibnamefont{Oogane}},
  \bibinfo{author}{\bibfnamefont{Y.}~\bibnamefont{Ando}},
  \bibinfo{author}{\bibfnamefont{H.~C.} \bibnamefont{Schneider}},
  \bibinfo{author}{\bibfnamefont{M.}~\bibnamefont{Aeschlimann}},
  \bibnamefont{and}
  \bibinfo{author}{\bibfnamefont{M.}~\bibnamefont{Cinchetti}},
  \bibinfo{journal}{Phys. Rev. Lett.} \textbf{\bibinfo{volume}{105}},
  \bibinfo{pages}{217202} (\bibinfo{year}{2010}),
  \urlprefix\url{https://link.aps.org/doi/10.1103/PhysRevLett.105.217202}.

\bibitem[{\citenamefont{Liu et~al.}(2010)\citenamefont{Liu, Shelford, Kruglyak,
  Hicken, Sakuraba, Oogane, and Ando}}]{Liu:2010cr}
\bibinfo{author}{\bibfnamefont{Y.}~\bibnamefont{Liu}},
  \bibinfo{author}{\bibfnamefont{L.~R.} \bibnamefont{Shelford}},
  \bibinfo{author}{\bibfnamefont{V.~V.} \bibnamefont{Kruglyak}},
  \bibinfo{author}{\bibfnamefont{R.~J.} \bibnamefont{Hicken}},
  \bibinfo{author}{\bibfnamefont{Y.}~\bibnamefont{Sakuraba}},
  \bibinfo{author}{\bibfnamefont{M.}~\bibnamefont{Oogane}}, \bibnamefont{and}
  \bibinfo{author}{\bibfnamefont{Y.}~\bibnamefont{Ando}},
  \bibinfo{journal}{Phys. Rev. B} \textbf{\bibinfo{volume}{81}},
  \bibinfo{pages}{094402} (\bibinfo{year}{2010}),
  \urlprefix\url{https://link.aps.org/doi/10.1103/PhysRevB.81.094402}.

\bibitem[{\citenamefont{Koopmans et~al.}(2009)\citenamefont{Koopmans,
  Malinowski, Longa, Steiauf, F{\"a}hnle, Roth, Cinchetti, and
  Aeschlimann}}]{Koopmans:2009ds}
\bibinfo{author}{\bibfnamefont{B.}~\bibnamefont{Koopmans}},
  \bibinfo{author}{\bibfnamefont{G.}~\bibnamefont{Malinowski}},
  \bibinfo{author}{\bibfnamefont{F.~D.} \bibnamefont{Longa}},
  \bibinfo{author}{\bibfnamefont{D.}~\bibnamefont{Steiauf}},
  \bibinfo{author}{\bibfnamefont{M.}~\bibnamefont{F{\"a}hnle}},
  \bibinfo{author}{\bibfnamefont{T.}~\bibnamefont{Roth}},
  \bibinfo{author}{\bibfnamefont{M.}~\bibnamefont{Cinchetti}},
  \bibnamefont{and}
  \bibinfo{author}{\bibfnamefont{M.}~\bibnamefont{Aeschlimann}},
  \bibinfo{journal}{Nature Publishing Group} \textbf{\bibinfo{volume}{9}},
  \bibinfo{pages}{259} (\bibinfo{year}{2009}),
  \urlprefix\url{https://www.nature.com/articles/nmat2593}.

\bibitem[{\citenamefont{Radu et~al.}(2011)\citenamefont{Radu, Vahaplar, Stamm,
  Kachel, Pontius, D{\"u}rr, Ostler, Barker, Evans, Chantrell
  et~al.}}]{Radu:2011kr}
\bibinfo{author}{\bibfnamefont{I.}~\bibnamefont{Radu}},
  \bibinfo{author}{\bibfnamefont{K.}~\bibnamefont{Vahaplar}},
  \bibinfo{author}{\bibfnamefont{C.}~\bibnamefont{Stamm}},
  \bibinfo{author}{\bibfnamefont{T.}~\bibnamefont{Kachel}},
  \bibinfo{author}{\bibfnamefont{N.}~\bibnamefont{Pontius}},
  \bibinfo{author}{\bibfnamefont{H.~A.} \bibnamefont{D{\"u}rr}},
  \bibinfo{author}{\bibfnamefont{T.~A.} \bibnamefont{Ostler}},
  \bibinfo{author}{\bibfnamefont{J.}~\bibnamefont{Barker}},
  \bibinfo{author}{\bibfnamefont{R.~F.~L.} \bibnamefont{Evans}},
  \bibinfo{author}{\bibfnamefont{R.~W.} \bibnamefont{Chantrell}},
  \bibnamefont{et~al.}, \bibinfo{journal}{Nature}
  \textbf{\bibinfo{volume}{472}}, \bibinfo{pages}{205} (\bibinfo{year}{2011}),
  \urlprefix\url{https://www.nature.com/articles/nature09901}.

\bibitem[{\citenamefont{Kazantseva et~al.}(2008)\citenamefont{Kazantseva,
  Nowak, Chantrell, Hohlfeld, and Rebei}}]{Kazantseva:2008bq}
\bibinfo{author}{\bibfnamefont{N.}~\bibnamefont{Kazantseva}},
  \bibinfo{author}{\bibfnamefont{U.}~\bibnamefont{Nowak}},
  \bibinfo{author}{\bibfnamefont{R.~W.} \bibnamefont{Chantrell}},
  \bibinfo{author}{\bibfnamefont{J.}~\bibnamefont{Hohlfeld}}, \bibnamefont{and}
  \bibinfo{author}{\bibfnamefont{A.}~\bibnamefont{Rebei}},
  \bibinfo{journal}{EPL} \textbf{\bibinfo{volume}{81}}, \bibinfo{pages}{27004}
  (\bibinfo{year}{2008}),
  \urlprefix\url{http://iopscience.iop.org/article/10.1209/0295-5075/81/27004}.

\bibitem[{\citenamefont{Atxitia et~al.}(2007)\citenamefont{Atxitia,
  Chubykalo-Fesenko, Kazantseva, Hinzke, Nowak, and
  Chantrell}}]{Atxitia:2007cq}
\bibinfo{author}{\bibfnamefont{U.}~\bibnamefont{Atxitia}},
  \bibinfo{author}{\bibfnamefont{O.}~\bibnamefont{Chubykalo-Fesenko}},
  \bibinfo{author}{\bibfnamefont{N.}~\bibnamefont{Kazantseva}},
  \bibinfo{author}{\bibfnamefont{D.}~\bibnamefont{Hinzke}},
  \bibinfo{author}{\bibfnamefont{U.}~\bibnamefont{Nowak}}, \bibnamefont{and}
  \bibinfo{author}{\bibfnamefont{R.~W.} \bibnamefont{Chantrell}},
  \bibinfo{journal}{Applied Physics Letters} \textbf{\bibinfo{volume}{91}},
  \bibinfo{pages}{232507} (\bibinfo{year}{2007}),
  \urlprefix\url{http://aip.scitation.org/doi/10.1063/1.2822807}.

\bibitem[{\citenamefont{Battiato et~al.}(2010)\citenamefont{Battiato, Carva,
  and Oppeneer}}]{Battiato:2010br}
\bibinfo{author}{\bibfnamefont{M.}~\bibnamefont{Battiato}},
  \bibinfo{author}{\bibfnamefont{K.}~\bibnamefont{Carva}}, \bibnamefont{and}
  \bibinfo{author}{\bibfnamefont{P.~M.} \bibnamefont{Oppeneer}},
  \bibinfo{journal}{Phys. Rev. Lett.} \textbf{\bibinfo{volume}{105}},
  \bibinfo{pages}{027203} (\bibinfo{year}{2010}),
  \urlprefix\url{https://link.aps.org/doi/10.1103/PhysRevLett.105.027203}.

\bibitem[{\citenamefont{Melnikov et~al.}(2011)\citenamefont{Melnikov,
  Razdolski, Wehling, Papaioannou, Roddatis, Fumagalli, Aktsipetrov,
  Lichtenstein, and Bovensiepen}}]{Melnikov:2011ep}
\bibinfo{author}{\bibfnamefont{A.}~\bibnamefont{Melnikov}},
  \bibinfo{author}{\bibfnamefont{I.}~\bibnamefont{Razdolski}},
  \bibinfo{author}{\bibfnamefont{T.~O.} \bibnamefont{Wehling}},
  \bibinfo{author}{\bibfnamefont{E.~T.} \bibnamefont{Papaioannou}},
  \bibinfo{author}{\bibfnamefont{V.}~\bibnamefont{Roddatis}},
  \bibinfo{author}{\bibfnamefont{P.}~\bibnamefont{Fumagalli}},
  \bibinfo{author}{\bibfnamefont{O.}~\bibnamefont{Aktsipetrov}},
  \bibinfo{author}{\bibfnamefont{A.~I.} \bibnamefont{Lichtenstein}},
  \bibnamefont{and}
  \bibinfo{author}{\bibfnamefont{U.}~\bibnamefont{Bovensiepen}},
  \bibinfo{journal}{Phys. Rev. Lett.} \textbf{\bibinfo{volume}{107}},
  \bibinfo{pages}{076601} (\bibinfo{year}{2011}),
  \urlprefix\url{https://link.aps.org/doi/10.1103/PhysRevLett.107.076601}.

\bibitem[{\citenamefont{Carva et~al.}(2011)\citenamefont{Carva, Battiato, and
  Oppeneer}}]{Carva:2011dp}
\bibinfo{author}{\bibfnamefont{K.}~\bibnamefont{Carva}},
  \bibinfo{author}{\bibfnamefont{M.}~\bibnamefont{Battiato}}, \bibnamefont{and}
  \bibinfo{author}{\bibfnamefont{P.~M.} \bibnamefont{Oppeneer}},
  \bibinfo{journal}{Phys. Rev. Lett.} \textbf{\bibinfo{volume}{107}},
  \bibinfo{pages}{207201} (\bibinfo{year}{2011}),
  \urlprefix\url{https://link.aps.org/doi/10.1103/PhysRevLett.107.207201}.

\bibitem[{\citenamefont{Essert and Schneider}(2011)}]{Essert:2011iq}
\bibinfo{author}{\bibfnamefont{S.}~\bibnamefont{Essert}} \bibnamefont{and}
  \bibinfo{author}{\bibfnamefont{H.~C.} \bibnamefont{Schneider}},
  \bibinfo{journal}{Phys. Rev. B} \textbf{\bibinfo{volume}{84}},
  \bibinfo{pages}{224405} (\bibinfo{year}{2011}),
  \urlprefix\url{https://link.aps.org/doi/10.1103/PhysRevB.84.224405}.

\bibitem[{\citenamefont{Atxitia and Chubykalo-Fesenko}(2011)}]{Atxitia:2011gn}
\bibinfo{author}{\bibfnamefont{U.}~\bibnamefont{Atxitia}} \bibnamefont{and}
  \bibinfo{author}{\bibfnamefont{O.}~\bibnamefont{Chubykalo-Fesenko}},
  \bibinfo{journal}{Phys. Rev. B} \textbf{\bibinfo{volume}{84}},
  \bibinfo{pages}{144414} (\bibinfo{year}{2011}),
  \urlprefix\url{https://link.aps.org/doi/10.1103/PhysRevB.84.144414}.

\bibitem[{\citenamefont{F{\"a}hnle and Illg}(2011)}]{Fahnle:2011eg}
\bibinfo{author}{\bibfnamefont{M.}~\bibnamefont{F{\"a}hnle}} \bibnamefont{and}
  \bibinfo{author}{\bibfnamefont{C.}~\bibnamefont{Illg}}, \bibinfo{journal}{J.
  Phys.: Condens. Matter} \textbf{\bibinfo{volume}{23}},
  \bibinfo{pages}{493201} (\bibinfo{year}{2011}),
  \urlprefix\url{http://iopscience.iop.org/article/10.1088/0953-8984/23/49/493201}.

\bibitem[{\citenamefont{Gilmore et~al.}(2007)\citenamefont{Gilmore, Idzerda,
  and Stiles}}]{Gilmore:2007evb}
\bibinfo{author}{\bibfnamefont{K.}~\bibnamefont{Gilmore}},
  \bibinfo{author}{\bibfnamefont{Y.~U.} \bibnamefont{Idzerda}},
  \bibnamefont{and} \bibinfo{author}{\bibfnamefont{M.~D.}
  \bibnamefont{Stiles}}, \bibinfo{journal}{Phys. Rev. Lett.}
  \textbf{\bibinfo{volume}{99}}, \bibinfo{pages}{027204}
  (\bibinfo{year}{2007}),
  \urlprefix\url{https://link.aps.org/doi/10.1103/PhysRevLett.99.027204}.

\bibitem[{\citenamefont{Brataas et~al.}(2008)\citenamefont{Brataas,
  Tserkovnyak, and Bauer}}]{Brataas:2008eqb}
\bibinfo{author}{\bibfnamefont{A.}~\bibnamefont{Brataas}},
  \bibinfo{author}{\bibfnamefont{Y.}~\bibnamefont{Tserkovnyak}},
  \bibnamefont{and} \bibinfo{author}{\bibfnamefont{G.~E.~W.}
  \bibnamefont{Bauer}}, \bibinfo{journal}{Phys. Rev. Lett.}
  \textbf{\bibinfo{volume}{101}}, \bibinfo{pages}{037207}
  (\bibinfo{year}{2008}),
  \urlprefix\url{https://link.aps.org/doi/10.1103/PhysRevLett.101.037207}.

\bibitem[{\citenamefont{Ebert et~al.}(2011{\natexlab{a}})\citenamefont{Ebert,
  Mankovsky, K{\"o}dderitzsch, and Kelly}}]{Ebert:2011gx}
\bibinfo{author}{\bibfnamefont{H.}~\bibnamefont{Ebert}},
  \bibinfo{author}{\bibfnamefont{S.}~\bibnamefont{Mankovsky}},
  \bibinfo{author}{\bibfnamefont{D.}~\bibnamefont{K{\"o}dderitzsch}},
  \bibnamefont{and} \bibinfo{author}{\bibfnamefont{P.~J.} \bibnamefont{Kelly}},
  \bibinfo{journal}{Phys. Rev. Lett.} \textbf{\bibinfo{volume}{107}},
  \bibinfo{pages}{066603} (\bibinfo{year}{2011}{\natexlab{a}}),
  \urlprefix\url{http://link.aps.org/doi/10.1103/PhysRevLett.107.066603}.

\bibitem[{\citenamefont{Itoh et~al.}(1983)\citenamefont{Itoh, Nakamichi,
  Yamaguchi, and Kazama}}]{Itoh:1983hy}
\bibinfo{author}{\bibfnamefont{H.}~\bibnamefont{Itoh}},
  \bibinfo{author}{\bibfnamefont{T.}~\bibnamefont{Nakamichi}},
  \bibinfo{author}{\bibfnamefont{Y.}~\bibnamefont{Yamaguchi}},
  \bibnamefont{and} \bibinfo{author}{\bibfnamefont{N.}~\bibnamefont{Kazama}},
  \bibinfo{journal}{Transactions of the Japan Institute of Metals}
  \textbf{\bibinfo{volume}{24}}, \bibinfo{pages}{265} (\bibinfo{year}{1983}),
  \urlprefix\url{https://www.jstage.jst.go.jp/article/matertrans1960/24/5/24_5_265/_article}.

\bibitem[{\citenamefont{Yutaka et~al.}(2013)\citenamefont{Yutaka, Masayuki, and
  Takuro}}]{Yutaka:2013ki}
\bibinfo{author}{\bibfnamefont{Y.}~\bibnamefont{Yutaka}},
  \bibinfo{author}{\bibfnamefont{K.}~\bibnamefont{Masayuki}}, \bibnamefont{and}
  \bibinfo{author}{\bibfnamefont{N.}~\bibnamefont{Takuro}},
  \bibinfo{journal}{J. Phys. Soc. Jpn.} \textbf{\bibinfo{volume}{50}},
  \bibinfo{pages}{2203} (\bibinfo{year}{2013}),
  \urlprefix\url{http://journals.jps.jp/doi/10.1143/JPSJ.50.2203}.

\bibitem[{\citenamefont{Jiang et~al.}(2001)\citenamefont{Jiang, Venkatesan, and
  Coey}}]{Jiang:2001kd}
\bibinfo{author}{\bibfnamefont{C.}~\bibnamefont{Jiang}},
  \bibinfo{author}{\bibfnamefont{M.}~\bibnamefont{Venkatesan}},
  \bibnamefont{and} \bibinfo{author}{\bibfnamefont{J.~M.~D.}
  \bibnamefont{Coey}}, \bibinfo{journal}{Solid State Communications}
  \textbf{\bibinfo{volume}{118}}, \bibinfo{pages}{513} (\bibinfo{year}{2001}),
  \urlprefix\url{http://linkinghub.elsevier.com/retrieve/pii/S003810980100151X}.

\bibitem[{\citenamefont{Shoji et~al.}(2013)\citenamefont{Shoji, Setsuro, and
  Junji}}]{Shoji:2013ge}
\bibinfo{author}{\bibfnamefont{I.}~\bibnamefont{Shoji}},
  \bibinfo{author}{\bibfnamefont{A.}~\bibnamefont{Setsuro}}, \bibnamefont{and}
  \bibinfo{author}{\bibfnamefont{I.}~\bibnamefont{Junji}}, \bibinfo{journal}{J.
  Phys. Soc. Jpn.} \textbf{\bibinfo{volume}{53}}, \bibinfo{pages}{2718}
  (\bibinfo{year}{2013}),
  \urlprefix\url{http://journals.jps.jp/doi/10.1143/JPSJ.53.2718}.

\bibitem[{\citenamefont{{\"O}zdogan et~al.}(2006)\citenamefont{{\"O}zdogan,
  Galanakis, {\c{S}}a{\c s}ioglu, and Akta{\c s}}}]{Ozdogan:2006jc}
\bibinfo{author}{\bibfnamefont{K.}~\bibnamefont{{\"O}zdogan}},
  \bibinfo{author}{\bibfnamefont{I.}~\bibnamefont{Galanakis}},
  \bibinfo{author}{\bibfnamefont{E.}~\bibnamefont{{\c{S}}a{\c s}ioglu}},
  \bibnamefont{and} \bibinfo{author}{\bibfnamefont{B.}~\bibnamefont{Akta{\c
  s}}}, \bibinfo{journal}{J. Phys.: Condens. Matter}
  \textbf{\bibinfo{volume}{18}}, \bibinfo{pages}{2905} (\bibinfo{year}{2006}),
  \urlprefix\url{http://iopscience.iop.org/article/10.1088/0953-8984/18/10/013}.

\bibitem[{\citenamefont{{\c{S}}a{\c s}ioglu
  et~al.}(2005)\citenamefont{{\c{S}}a{\c s}ioglu, Sandratskii, and
  Bruno}}]{Sasioglu:2005ex}
\bibinfo{author}{\bibfnamefont{E.}~\bibnamefont{{\c{S}}a{\c s}ioglu}},
  \bibinfo{author}{\bibfnamefont{L.~M.} \bibnamefont{Sandratskii}},
  \bibnamefont{and} \bibinfo{author}{\bibfnamefont{P.}~\bibnamefont{Bruno}},
  \bibinfo{journal}{J. Phys.: Condens. Matter} \textbf{\bibinfo{volume}{17}},
  \bibinfo{pages}{995} (\bibinfo{year}{2005}),
  \urlprefix\url{http://iopscience.iop.org/article/10.1088/0953-8984/17/6/017}.

\bibitem[{\citenamefont{Weht and Pickett}(1999)}]{Weht:1999gq}
\bibinfo{author}{\bibfnamefont{R.}~\bibnamefont{Weht}} \bibnamefont{and}
  \bibinfo{author}{\bibfnamefont{W.~E.} \bibnamefont{Pickett}},
  \bibinfo{journal}{Phys. Rev. B} \textbf{\bibinfo{volume}{60}},
  \bibinfo{pages}{13006} (\bibinfo{year}{1999}),
  \urlprefix\url{https://link.aps.org/doi/10.1103/PhysRevB.60.13006}.

\bibitem[{\citenamefont{Liu et~al.}(2008)\citenamefont{Liu, Dai, Liu, Chen, Li,
  Xiao, and Wu}}]{Liu:2008dt}
\bibinfo{author}{\bibfnamefont{G.~D.} \bibnamefont{Liu}},
  \bibinfo{author}{\bibfnamefont{X.~F.} \bibnamefont{Dai}},
  \bibinfo{author}{\bibfnamefont{H.~Y.} \bibnamefont{Liu}},
  \bibinfo{author}{\bibfnamefont{J.~L.} \bibnamefont{Chen}},
  \bibinfo{author}{\bibfnamefont{Y.~X.} \bibnamefont{Li}},
  \bibinfo{author}{\bibfnamefont{G.}~\bibnamefont{Xiao}}, \bibnamefont{and}
  \bibinfo{author}{\bibfnamefont{G.~H.} \bibnamefont{Wu}},
  \bibinfo{journal}{Phys. Rev. B} \textbf{\bibinfo{volume}{77}},
  \bibinfo{pages}{014424} (\bibinfo{year}{2008}),
  \urlprefix\url{https://link.aps.org/doi/10.1103/PhysRevB.77.014424}.

\bibitem[{\citenamefont{Ouardi et~al.}(2013)\citenamefont{Ouardi, Fecher,
  Felser, and K{\"u}bler}}]{Ouardi:2013ko}
\bibinfo{author}{\bibfnamefont{S.}~\bibnamefont{Ouardi}},
  \bibinfo{author}{\bibfnamefont{G.~H.} \bibnamefont{Fecher}},
  \bibinfo{author}{\bibfnamefont{C.}~\bibnamefont{Felser}}, \bibnamefont{and}
  \bibinfo{author}{\bibfnamefont{J.}~\bibnamefont{K{\"u}bler}},
  \bibinfo{journal}{Phys. Rev. Lett.} \textbf{\bibinfo{volume}{110}},
  \bibinfo{pages}{100401} (\bibinfo{year}{2013}),
  \urlprefix\url{https://link.aps.org/doi/10.1103/PhysRevLett.110.100401}.

\bibitem[{\citenamefont{Zabloudil et~al.}(2006)\citenamefont{Zabloudil,
  Szunyogh, Hammerling, and Weinberger}}]{Zabloudil:2005dJ}
\bibinfo{author}{\bibfnamefont{J.}~\bibnamefont{Zabloudil}},
  \bibinfo{author}{\bibfnamefont{L.}~\bibnamefont{Szunyogh}},
  \bibinfo{author}{\bibfnamefont{R.}~\bibnamefont{Hammerling}},
  \bibnamefont{and}
  \bibinfo{author}{\bibfnamefont{P.}~\bibnamefont{Weinberger}},
  \emph{\bibinfo{title}{{Electron Scattering in Solid Matter}}}, A Theoretical
  and Computational Treatise (\bibinfo{year}{2006}),
  \urlprefix\url{http://bookzz.org/md5/190C5DE184B30E2B6898DE499DFB7D78}.

\bibitem[{\citenamefont{Ebert et~al.}(2011{\natexlab{b}})\citenamefont{Ebert,
  K{\"o}dderitzsch, and Min{\'a}r}}]{Ebert:2011di}
\bibinfo{author}{\bibfnamefont{H.}~\bibnamefont{Ebert}},
  \bibinfo{author}{\bibfnamefont{D.}~\bibnamefont{K{\"o}dderitzsch}},
  \bibnamefont{and}
  \bibinfo{author}{\bibfnamefont{J.}~\bibnamefont{Min{\'a}r}},
  \bibinfo{journal}{Rep. Prog. Phys.} \textbf{\bibinfo{volume}{74}},
  \bibinfo{pages}{096501} (\bibinfo{year}{2011}{\natexlab{b}}),
  \urlprefix\url{http://iopscience.iop.org/article/10.1088/0034-4885/74/9/096501}.

\bibitem[{\citenamefont{Ebert}(2012)}]{Ebert:PzMUDULG}
\bibinfo{author}{\bibfnamefont{H.}~\bibnamefont{Ebert}},
  \emph{\bibinfo{title}{{The Munich SPR-KKR package, version 6.3,}}}
  (\bibinfo{year}{2012}),
  \urlprefix\url{http://ebert.cup.uni-muenchen.de/SPRKKR}.

\bibitem[{\citenamefont{Gross and Dreizler}(2013)}]{Gross:2013jq}
\bibinfo{author}{\bibfnamefont{E.~K.~U.} \bibnamefont{Gross}} \bibnamefont{and}
  \bibinfo{author}{\bibfnamefont{R.~M.} \bibnamefont{Dreizler}},
  \emph{\bibinfo{title}{{Density Functional Theory}}}
  (\bibinfo{publisher}{Springer Science {\&} Business Media},
  \bibinfo{year}{2013}), ISBN \bibinfo{isbn}{1475799756},
  \urlprefix\url{http://books.google.se/books?id=aG4ECAAAQBAJ&pg=PR4&dq=10.1007/978-1-4757-9975-0&hl=&cd=1&source=gbs_api}.

\bibitem[{\citenamefont{Perdew et~al.}(1996)\citenamefont{Perdew, Burke, and
  Ernzerhof}}]{Perdew:1996iq}
\bibinfo{author}{\bibfnamefont{J.~P.} \bibnamefont{Perdew}},
  \bibinfo{author}{\bibfnamefont{K.}~\bibnamefont{Burke}}, \bibnamefont{and}
  \bibinfo{author}{\bibfnamefont{M.}~\bibnamefont{Ernzerhof}},
  \bibinfo{journal}{Phys. Rev. Lett.} \textbf{\bibinfo{volume}{77}},
  \bibinfo{pages}{3865} (\bibinfo{year}{1996}),
  \urlprefix\url{https://link.aps.org/doi/10.1103/PhysRevLett.77.3865}.

\bibitem[{\citenamefont{Gyorffy}(1972)}]{Gyorffy:1972df}
\bibinfo{author}{\bibfnamefont{B.~L.} \bibnamefont{Gyorffy}},
  \bibinfo{journal}{Phys. Rev. B} \textbf{\bibinfo{volume}{5}},
  \bibinfo{pages}{2382} (\bibinfo{year}{1972}),
  \urlprefix\url{https://link.aps.org/doi/10.1103/PhysRevB.5.2382}.

\bibitem[{\citenamefont{Huhne et~al.}(1998)\citenamefont{Huhne, Zecha, Ebert,
  Dederichs, and Zeller}}]{1998PhRvB..5810236H}
\bibinfo{author}{\bibfnamefont{T.}~\bibnamefont{Huhne}},
  \bibinfo{author}{\bibfnamefont{C.}~\bibnamefont{Zecha}},
  \bibinfo{author}{\bibfnamefont{H.}~\bibnamefont{Ebert}},
  \bibinfo{author}{\bibfnamefont{P.~H.} \bibnamefont{Dederichs}},
  \bibnamefont{and} \bibinfo{author}{\bibfnamefont{R.}~\bibnamefont{Zeller}},
  \bibinfo{journal}{Physical Review B (Condensed Matter and Materials Physics)}
  \textbf{\bibinfo{volume}{58}}, \bibinfo{pages}{10236} (\bibinfo{year}{1998}),
  \urlprefix\url{http://adsabs.harvard.edu/cgi-bin/nph-data_query?bibcode=1998PhRvB..5810236H&link_type=EJOURNAL}.

\bibitem[{\citenamefont{Liechtenstein et~al.}(1984)\citenamefont{Liechtenstein,
  Katsnelson, and Gubanov}}]{Liechtenstein:1984fj}
\bibinfo{author}{\bibfnamefont{A.~I.} \bibnamefont{Liechtenstein}},
  \bibinfo{author}{\bibfnamefont{M.~I.} \bibnamefont{Katsnelson}},
  \bibnamefont{and} \bibinfo{author}{\bibfnamefont{V.~A.}
  \bibnamefont{Gubanov}}, \bibinfo{journal}{J. Phys. F: Met. Phys.}
  \textbf{\bibinfo{volume}{14}}, \bibinfo{pages}{L125} (\bibinfo{year}{1984}),
  \urlprefix\url{http://iopscience.iop.org/article/10.1088/0305-4608/14/7/007}.

\bibitem[{\citenamefont{Pajda et~al.}(2001)\citenamefont{Pajda,
  Kudrnovsk{\'{y}}, Turek, Drchal, and Bruno}}]{Pajda:2001ix}
\bibinfo{author}{\bibfnamefont{M.}~\bibnamefont{Pajda}},
  \bibinfo{author}{\bibfnamefont{J.}~\bibnamefont{Kudrnovsk{\'{y}}}},
  \bibinfo{author}{\bibfnamefont{I.}~\bibnamefont{Turek}},
  \bibinfo{author}{\bibfnamefont{V.}~\bibnamefont{Drchal}}, \bibnamefont{and}
  \bibinfo{author}{\bibfnamefont{P.}~\bibnamefont{Bruno}},
  \bibinfo{journal}{Phys. Rev. B} \textbf{\bibinfo{volume}{64}},
  \bibinfo{pages}{174402} (\bibinfo{year}{2001}),
  \urlprefix\url{https://link.aps.org/doi/10.1103/PhysRevB.64.174402}.

\bibitem[{\citenamefont{Mankovsky et~al.}(2013)\citenamefont{Mankovsky,
  K{\"o}dderitzsch, Woltersdorf, and Ebert}}]{Mankovsky:2013ii}
\bibinfo{author}{\bibfnamefont{S.}~\bibnamefont{Mankovsky}},
  \bibinfo{author}{\bibfnamefont{D.}~\bibnamefont{K{\"o}dderitzsch}},
  \bibinfo{author}{\bibfnamefont{G.}~\bibnamefont{Woltersdorf}},
  \bibnamefont{and} \bibinfo{author}{\bibfnamefont{H.}~\bibnamefont{Ebert}},
  \bibinfo{journal}{Phys. Rev. B} \textbf{\bibinfo{volume}{87}},
  \bibinfo{pages}{014430} (\bibinfo{year}{2013}),
  \urlprefix\url{http://link.aps.org/doi/10.1103/PhysRevB.87.014430}.

\bibitem[{\citenamefont{Butler}(1985)}]{Butler:1985by}
\bibinfo{author}{\bibfnamefont{W.~H.} \bibnamefont{Butler}},
  \bibinfo{journal}{Phys. Rev. B} \textbf{\bibinfo{volume}{31}},
  \bibinfo{pages}{3260} (\bibinfo{year}{1985}),
  \urlprefix\url{https://link.aps.org/doi/10.1103/PhysRevB.31.3260}.

\bibitem[{\citenamefont{Ebert et~al.}(2015)\citenamefont{Ebert, Mankovsky,
  Chadova, Polesya, Min{\'a}r, and K{\"o}dderitzsch}}]{Ebert:2015kxa}
\bibinfo{author}{\bibfnamefont{H.}~\bibnamefont{Ebert}},
  \bibinfo{author}{\bibfnamefont{S.}~\bibnamefont{Mankovsky}},
  \bibinfo{author}{\bibfnamefont{K.}~\bibnamefont{Chadova}},
  \bibinfo{author}{\bibfnamefont{S.}~\bibnamefont{Polesya}},
  \bibinfo{author}{\bibfnamefont{J.}~\bibnamefont{Min{\'a}r}},
  \bibnamefont{and}
  \bibinfo{author}{\bibfnamefont{D.}~\bibnamefont{K{\"o}dderitzsch}},
  \bibinfo{journal}{Phys. Rev. B} \textbf{\bibinfo{volume}{91}},
  \bibinfo{pages}{165132} (\bibinfo{year}{2015}),
  \urlprefix\url{https://link.aps.org/doi/10.1103/PhysRevB.91.165132}.

\bibitem[{\citenamefont{Antropov et~al.}(1996)\citenamefont{Antropov,
  Katsnelson, Harmon, van Schilfgaarde, and Kusnezov}}]{Antropov:1996td}
\bibinfo{author}{\bibfnamefont{V.~P.} \bibnamefont{Antropov}},
  \bibinfo{author}{\bibfnamefont{M.~I.} \bibnamefont{Katsnelson}},
  \bibinfo{author}{\bibfnamefont{B.~N.} \bibnamefont{Harmon}},
  \bibinfo{author}{\bibfnamefont{M.}~\bibnamefont{van Schilfgaarde}},
  \bibnamefont{and} \bibinfo{author}{\bibfnamefont{D.}~\bibnamefont{Kusnezov}},
  \bibinfo{journal}{Phys. Rev. B} \textbf{\bibinfo{volume}{54}},
  \bibinfo{pages}{1019} (\bibinfo{year}{1996}),
  \urlprefix\url{http://link.aps.org/doi/10.1103/PhysRevB.54.1019}.

\bibitem[{\citenamefont{Eriksson et~al.}(2016)\citenamefont{Eriksson, Bergman,
  Bergqvist, and Hellsvik}}]{Eriksson:2016uw}
\bibinfo{author}{\bibfnamefont{O.}~\bibnamefont{Eriksson}},
  \bibinfo{author}{\bibfnamefont{A.}~\bibnamefont{Bergman}},
  \bibinfo{author}{\bibfnamefont{L.}~\bibnamefont{Bergqvist}},
  \bibnamefont{and} \bibinfo{author}{\bibfnamefont{J.}~\bibnamefont{Hellsvik}},
  \emph{\bibinfo{title}{{Atomistic Spin Dynamics}}}, Foundations and
  Applications (\bibinfo{publisher}{Oxford University Press},
  \bibinfo{year}{2016}),
  \urlprefix\url{https://global.oup.com/academic/product/atomistic-spin-dynamics-9780198788669}.

\bibitem[{\citenamefont{Bovensiepen}(2007)}]{Bovensiepen:2007gc}
\bibinfo{author}{\bibfnamefont{U.}~\bibnamefont{Bovensiepen}},
  \bibinfo{journal}{J. Phys.: Condens. Matter} \textbf{\bibinfo{volume}{19}},
  \bibinfo{pages}{083201} (\bibinfo{year}{2007}),
  \urlprefix\url{http://iopscience.iop.org/article/10.1088/0953-8984/19/8/083201}.

\bibitem[{\citenamefont{Chimata et~al.}(2012)\citenamefont{Chimata, Bergman,
  Bergqvist, Sanyal, and Eriksson}}]{Chimata:2012kv}
\bibinfo{author}{\bibfnamefont{R.}~\bibnamefont{Chimata}},
  \bibinfo{author}{\bibfnamefont{A.}~\bibnamefont{Bergman}},
  \bibinfo{author}{\bibfnamefont{L.}~\bibnamefont{Bergqvist}},
  \bibinfo{author}{\bibfnamefont{B.}~\bibnamefont{Sanyal}}, \bibnamefont{and}
  \bibinfo{author}{\bibfnamefont{O.}~\bibnamefont{Eriksson}},
  \bibinfo{journal}{Phys. Rev. Lett.} \textbf{\bibinfo{volume}{109}},
  \bibinfo{pages}{157201} (\bibinfo{year}{2012}),
  \urlprefix\url{https://link.aps.org/doi/10.1103/PhysRevLett.109.157201}.

\bibitem[{\citenamefont{Jakobsson et~al.}(2015)\citenamefont{Jakobsson,
  Mavropoulos, {\c{S}}a{\c s}{\i}o{\u{g}}lu, Bl{\"u}gel, Le{\v z}ai{\'c},
  Sanyal, and Galanakis}}]{Jakobsson:2015dt}
\bibinfo{author}{\bibfnamefont{A.}~\bibnamefont{Jakobsson}},
  \bibinfo{author}{\bibfnamefont{P.}~\bibnamefont{Mavropoulos}},
  \bibinfo{author}{\bibfnamefont{E.}~\bibnamefont{{\c{S}}a{\c
  s}{\i}o{\u{g}}lu}},
  \bibinfo{author}{\bibfnamefont{S.}~\bibnamefont{Bl{\"u}gel}},
  \bibinfo{author}{\bibfnamefont{M.}~\bibnamefont{Le{\v z}ai{\'c}}},
  \bibinfo{author}{\bibfnamefont{B.}~\bibnamefont{Sanyal}}, \bibnamefont{and}
  \bibinfo{author}{\bibfnamefont{I.}~\bibnamefont{Galanakis}},
  \bibinfo{journal}{Phys. Rev. B} \textbf{\bibinfo{volume}{91}},
  \bibinfo{pages}{174439} (\bibinfo{year}{2015}),
  \urlprefix\url{http://link.aps.org/doi/10.1103/PhysRevB.91.174439}.

\bibitem[{\citenamefont{Motizuki et~al.}(2009)\citenamefont{Motizuki, Ido,
  Itoh, and Morifuji}}]{Motizuki:2009vs}
\bibinfo{author}{\bibfnamefont{K.}~\bibnamefont{Motizuki}},
  \bibinfo{author}{\bibfnamefont{H.}~\bibnamefont{Ido}},
  \bibinfo{author}{\bibfnamefont{T.}~\bibnamefont{Itoh}}, \bibnamefont{and}
  \bibinfo{author}{\bibfnamefont{M.}~\bibnamefont{Morifuji}},
  \emph{\bibinfo{title}{{Electronic Structure and Magnetism of 3d-Transition
  Metal Pnictides}}} (\bibinfo{publisher}{Springer Science {\&} Business
  Media}, \bibinfo{year}{2009}), ISBN \bibinfo{isbn}{3642034209},
  \urlprefix\url{http://books.google.se/books?id=g1wv4vHY58cC&printsec=frontcover&dq=intitle:Electronic+Structure+and+Magnetism+of+3d+Transition+Metal+Kazuko+Motizuki+Springer&hl=&cd=1&source=gbs_api}.

\bibitem[{\citenamefont{Chimata et~al.}(2017)\citenamefont{Chimata,
  Delczeg-Czirjak, Szilva, Cardias, Kvashnin, Pereiro, Mankovsky, Ebert,
  Thonig, Sanyal et~al.}}]{Chimata:2017iu}
\bibinfo{author}{\bibfnamefont{R.}~\bibnamefont{Chimata}},
  \bibinfo{author}{\bibfnamefont{E.~K.} \bibnamefont{Delczeg-Czirjak}},
  \bibinfo{author}{\bibfnamefont{A.}~\bibnamefont{Szilva}},
  \bibinfo{author}{\bibfnamefont{R.}~\bibnamefont{Cardias}},
  \bibinfo{author}{\bibfnamefont{Y.~O.} \bibnamefont{Kvashnin}},
  \bibinfo{author}{\bibfnamefont{M.}~\bibnamefont{Pereiro}},
  \bibinfo{author}{\bibfnamefont{S.}~\bibnamefont{Mankovsky}},
  \bibinfo{author}{\bibfnamefont{H.}~\bibnamefont{Ebert}},
  \bibinfo{author}{\bibfnamefont{D.}~\bibnamefont{Thonig}},
  \bibinfo{author}{\bibfnamefont{B.}~\bibnamefont{Sanyal}},
  \bibnamefont{et~al.}, \bibinfo{journal}{Phys. Rev. B}
  \textbf{\bibinfo{volume}{95}}, \bibinfo{pages}{214417}
  (\bibinfo{year}{2017}),
  \urlprefix\url{http://link.aps.org/doi/10.1103/PhysRevB.95.214417}.

\bibitem[{\citenamefont{Meinert et~al.}(2011)\citenamefont{Meinert,
  Schmalhorst, and Reiss}}]{Meinert:2011hn}
\bibinfo{author}{\bibfnamefont{M.}~\bibnamefont{Meinert}},
  \bibinfo{author}{\bibfnamefont{J.-M.} \bibnamefont{Schmalhorst}},
  \bibnamefont{and} \bibinfo{author}{\bibfnamefont{G.}~\bibnamefont{Reiss}},
  \bibinfo{journal}{J. Phys.: Condens. Matter} \textbf{\bibinfo{volume}{23}},
  \bibinfo{pages}{116005} (\bibinfo{year}{2011}),
  \urlprefix\url{http://iopscience.iop.org/article/10.1088/0953-8984/23/11/116005}.

\bibitem[{\citenamefont{Jiles}(2015)}]{Jiles:2015uu}
\bibinfo{author}{\bibfnamefont{D.}~\bibnamefont{Jiles}},
  \emph{\bibinfo{title}{{Introduction to Magnetism and Magnetic Materials,
  Third Edition}}} (\bibinfo{publisher}{CRC Press}, \bibinfo{year}{2015}), ISBN
  \bibinfo{isbn}{1482238888},
  \urlprefix\url{http://books.google.se/books?id=2diYCgAAQBAJ&printsec=frontcover&dq=intitle:Introduction+to+Magnetism+and+Magnetic+Materials+Second+Edition&hl=&cd=1&source=gbs_api}.

\bibitem[{\citenamefont{Rusz et~al.}(2006)\citenamefont{Rusz, Bergqvist,
  Kudrnovsk{\'{y}}, and Turek}}]{Rusz:2006jy}
\bibinfo{author}{\bibfnamefont{J.}~\bibnamefont{Rusz}},
  \bibinfo{author}{\bibfnamefont{L.}~\bibnamefont{Bergqvist}},
  \bibinfo{author}{\bibfnamefont{J.}~\bibnamefont{Kudrnovsk{\'{y}}}},
  \bibnamefont{and} \bibinfo{author}{\bibfnamefont{I.}~\bibnamefont{Turek}},
  \bibinfo{journal}{Phys. Rev. B} \textbf{\bibinfo{volume}{73}},
  \bibinfo{pages}{214412} (\bibinfo{year}{2006}),
  \urlprefix\url{https://link.aps.org/doi/10.1103/PhysRevB.73.214412}.

\bibitem[{\citenamefont{B{\"o}ttcher et~al.}(2012)\citenamefont{B{\"o}ttcher,
  Ernst, and Henk}}]{Bottcher:2012hz}
\bibinfo{author}{\bibfnamefont{D.}~\bibnamefont{B{\"o}ttcher}},
  \bibinfo{author}{\bibfnamefont{A.}~\bibnamefont{Ernst}}, \bibnamefont{and}
  \bibinfo{author}{\bibfnamefont{J.}~\bibnamefont{Henk}},
  \bibinfo{journal}{Journal of Magnetism and Magnetic Materials}
  \textbf{\bibinfo{volume}{324}}, \bibinfo{pages}{610} (\bibinfo{year}{2012}),
  \urlprefix\url{http://linkinghub.elsevier.com/retrieve/pii/S0304885311006299}.

\bibitem[{\citenamefont{Chico et~al.}(2016)\citenamefont{Chico, Keshavarz,
  Kvashnin, Pereiro, Di~Marco, Etz, Eriksson, Bergman, and
  Bergqvist}}]{Chico:2016dya}
\bibinfo{author}{\bibfnamefont{J.}~\bibnamefont{Chico}},
  \bibinfo{author}{\bibfnamefont{S.}~\bibnamefont{Keshavarz}},
  \bibinfo{author}{\bibfnamefont{Y.}~\bibnamefont{Kvashnin}},
  \bibinfo{author}{\bibfnamefont{M.}~\bibnamefont{Pereiro}},
  \bibinfo{author}{\bibfnamefont{I.}~\bibnamefont{Di~Marco}},
  \bibinfo{author}{\bibfnamefont{C.}~\bibnamefont{Etz}},
  \bibinfo{author}{\bibfnamefont{O.}~\bibnamefont{Eriksson}},
  \bibinfo{author}{\bibfnamefont{A.}~\bibnamefont{Bergman}}, \bibnamefont{and}
  \bibinfo{author}{\bibfnamefont{L.}~\bibnamefont{Bergqvist}},
  \bibinfo{journal}{Phys. Rev. B} \textbf{\bibinfo{volume}{93}},
  \bibinfo{pages}{214439} (\bibinfo{year}{2016}),
  \urlprefix\url{https://link.aps.org/doi/10.1103/PhysRevB.93.214439}.

\bibitem[{\citenamefont{Umetsu and Kanomata}(2015)}]{Umetsu:2015db}
\bibinfo{author}{\bibfnamefont{R.~Y.} \bibnamefont{Umetsu}} \bibnamefont{and}
  \bibinfo{author}{\bibfnamefont{T.}~\bibnamefont{Kanomata}},
  \bibinfo{journal}{Physics Procedia} \textbf{\bibinfo{volume}{75}},
  \bibinfo{pages}{890} (\bibinfo{year}{2015}),
  \urlprefix\url{http://linkinghub.elsevier.com/retrieve/pii/S187538921501754X}.

\bibitem[{\citenamefont{Binder and Heermann}(2010)}]{Anonymous:ypogcH9U}
\bibinfo{author}{\bibfnamefont{K.}~\bibnamefont{Binder}} \bibnamefont{and}
  \bibinfo{author}{\bibfnamefont{D.}~\bibnamefont{Heermann}},
  \emph{\bibinfo{title}{{Monte Carlo Simulation in Statistical Physics }}},
  vol.~\bibinfo{volume}{5} of \emph{\bibinfo{series}{An Introduction}}
  (\bibinfo{address}{Berlin Heidelberg}, \bibinfo{year}{2010}),
  \bibinfo{edition}{springer-verlag} ed.,
  \urlprefix\url{//www.springer.com/de/book/9783642031625}.

\bibitem[{\citenamefont{Meinel et~al.}(1995)\citenamefont{Meinel, Beckmann,
  Klaua, and Bethge}}]{Meinel:1995ct}
\bibinfo{author}{\bibfnamefont{K.}~\bibnamefont{Meinel}},
  \bibinfo{author}{\bibfnamefont{A.}~\bibnamefont{Beckmann}},
  \bibinfo{author}{\bibfnamefont{M.}~\bibnamefont{Klaua}}, \bibnamefont{and}
  \bibinfo{author}{\bibfnamefont{H.}~\bibnamefont{Bethge}},
  \bibinfo{journal}{physica status solidi (a)} \textbf{\bibinfo{volume}{150}},
  \bibinfo{pages}{521} (\bibinfo{year}{1995}),
  \urlprefix\url{http://doi.wiley.com/10.1002/pssa.2211500146}.

\bibitem[{\citenamefont{Staunton}(2013)}]{Staunton:2013ww}
\bibinfo{author}{\bibfnamefont{J.}~\bibnamefont{Staunton}},
  \emph{\bibinfo{title}{{Relativistic Effects and Disordered Local Moments in
  Magnets }}} (\bibinfo{year}{2013}),
  \urlprefix\url{http://www.psi-k.org/newsletters/News_82/Highlight_82.pdf}.

\bibitem[{\citenamefont{Akai and Dederichs}(1993)}]{Akai:1993jy}
\bibinfo{author}{\bibfnamefont{H.}~\bibnamefont{Akai}} \bibnamefont{and}
  \bibinfo{author}{\bibfnamefont{P.~H.} \bibnamefont{Dederichs}},
  \bibinfo{journal}{Phys. Rev. B} \textbf{\bibinfo{volume}{47}},
  \bibinfo{pages}{8739} (\bibinfo{year}{1993}),
  \urlprefix\url{https://link.aps.org/doi/10.1103/PhysRevB.47.8739}.

\bibitem[{\citenamefont{Buruzs}(2008)}]{Buruzs:Pim91Guq}
\bibinfo{author}{\bibfnamefont{A.}~\bibnamefont{Buruzs}}, Ph.D. thesis,
  \bibinfo{school}{cms.tuwien.ac.at}, \bibinfo{address}{Wien}
  (\bibinfo{year}{2008}),
  \urlprefix\url{http://www.cms.tuwien.ac.at/media/pdf/phd-thesis/THESIS.PDF}.

\bibitem[{\citenamefont{Schoen et~al.}(2016)\citenamefont{Schoen, Thonig,
  Schneider, Silva, Nembach, Eriksson, Karis, and Shaw}}]{Schoen:2016gcc}
\bibinfo{author}{\bibfnamefont{M.~A.~W.} \bibnamefont{Schoen}},
  \bibinfo{author}{\bibfnamefont{D.}~\bibnamefont{Thonig}},
  \bibinfo{author}{\bibfnamefont{M.~L.} \bibnamefont{Schneider}},
  \bibinfo{author}{\bibfnamefont{T.~J.} \bibnamefont{Silva}},
  \bibinfo{author}{\bibfnamefont{H.~T.} \bibnamefont{Nembach}},
  \bibinfo{author}{\bibfnamefont{O.}~\bibnamefont{Eriksson}},
  \bibinfo{author}{\bibfnamefont{O.}~\bibnamefont{Karis}}, \bibnamefont{and}
  \bibinfo{author}{\bibfnamefont{J.~M.} \bibnamefont{Shaw}},
  \bibinfo{journal}{Nat Phys} \textbf{\bibinfo{volume}{12}},
  \bibinfo{pages}{839} (\bibinfo{year}{2016}),
  \urlprefix\url{http://www.nature.com/doifinder/10.1038/nphys3770}.

\bibitem[{\citenamefont{Lounis et~al.}(2015)\citenamefont{Lounis, dos
  Santos~Dias, and Schweflinghaus}}]{Lounis:2015ho}
\bibinfo{author}{\bibfnamefont{S.}~\bibnamefont{Lounis}},
  \bibinfo{author}{\bibfnamefont{M.}~\bibnamefont{dos Santos~Dias}},
  \bibnamefont{and}
  \bibinfo{author}{\bibfnamefont{B.}~\bibnamefont{Schweflinghaus}},
  \bibinfo{journal}{Phys. Rev. B} \textbf{\bibinfo{volume}{91}},
  \bibinfo{pages}{104420} (\bibinfo{year}{2015}),
  \urlprefix\url{https://link.aps.org/doi/10.1103/PhysRevB.91.104420}.

\bibitem[{\citenamefont{Pan et~al.}(2016)\citenamefont{Pan, Chico, Hellsvik,
  Delin, Bergman, and Bergqvist}}]{Pan:2016gb}
\bibinfo{author}{\bibfnamefont{F.}~\bibnamefont{Pan}},
  \bibinfo{author}{\bibfnamefont{J.}~\bibnamefont{Chico}},
  \bibinfo{author}{\bibfnamefont{J.}~\bibnamefont{Hellsvik}},
  \bibinfo{author}{\bibfnamefont{A.}~\bibnamefont{Delin}},
  \bibinfo{author}{\bibfnamefont{A.}~\bibnamefont{Bergman}}, \bibnamefont{and}
  \bibinfo{author}{\bibfnamefont{L.}~\bibnamefont{Bergqvist}},
  \bibinfo{journal}{Phys. Rev. B} \textbf{\bibinfo{volume}{94}},
  \bibinfo{pages}{214410} (\bibinfo{year}{2016}),
  \urlprefix\url{https://link.aps.org/doi/10.1103/PhysRevB.94.214410}.

\bibitem[{\citenamefont{Kambersk{\'{y}}}(1984)}]{Kambersky:1984iz}
\bibinfo{author}{\bibfnamefont{V.}~\bibnamefont{Kambersk{\'{y}}}},
  \bibinfo{journal}{Czech J Phys} \textbf{\bibinfo{volume}{34}},
  \bibinfo{pages}{1111} (\bibinfo{year}{1984}),
  \urlprefix\url{https://link.springer.com/article/10.1007/BF01590106}.

\bibitem[{\citenamefont{Kambersk{\'{y}}}(1976)}]{Kambersky:1976gi}
\bibinfo{author}{\bibfnamefont{V.}~\bibnamefont{Kambersk{\'{y}}}},
  \bibinfo{journal}{Czech J Phys} \textbf{\bibinfo{volume}{26}},
  \bibinfo{pages}{1366} (\bibinfo{year}{1976}),
  \urlprefix\url{http://link.springer.com/10.1007/BF01587621}.

\bibitem[{\citenamefont{Skubic et~al.}(2008)\citenamefont{Skubic, Hellsvik,
  Nordstr{\"o}m, and Eriksson}}]{Skubic:2008gs}
\bibinfo{author}{\bibfnamefont{B.}~\bibnamefont{Skubic}},
  \bibinfo{author}{\bibfnamefont{J.}~\bibnamefont{Hellsvik}},
  \bibinfo{author}{\bibfnamefont{L.}~\bibnamefont{Nordstr{\"o}m}},
  \bibnamefont{and} \bibinfo{author}{\bibfnamefont{O.}~\bibnamefont{Eriksson}},
  \bibinfo{journal}{J. Phys.: Condens. Matter} \textbf{\bibinfo{volume}{20}},
  \bibinfo{pages}{315203} (\bibinfo{year}{2008}),
  \urlprefix\url{http://iopscience.iop.org/article/10.1088/0953-8984/20/31/315203}.

\bibitem[{\citenamefont{Berritta et~al.}(2016)\citenamefont{Berritta, Mondal,
  Carva, and Oppeneer}}]{Berritta:2016em}
\bibinfo{author}{\bibfnamefont{M.}~\bibnamefont{Berritta}},
  \bibinfo{author}{\bibfnamefont{R.}~\bibnamefont{Mondal}},
  \bibinfo{author}{\bibfnamefont{K.}~\bibnamefont{Carva}}, \bibnamefont{and}
  \bibinfo{author}{\bibfnamefont{P.~M.} \bibnamefont{Oppeneer}},
  \bibinfo{journal}{Phys. Rev. Lett.} \textbf{\bibinfo{volume}{117}},
  \bibinfo{pages}{137203} (\bibinfo{year}{2016}),
  \urlprefix\url{https://link.aps.org/doi/10.1103/PhysRevLett.117.137203}.

\bibitem[{\citenamefont{John et~al.}(2017)\citenamefont{John, Berritta, Hinzke,
  M{\"u}ller, Santos, Ulrichs, Nieves, Walowski, Mondal, Chubykalo-Fesenko
  et~al.}}]{2017NatSR...7.4114J}
\bibinfo{author}{\bibfnamefont{R.}~\bibnamefont{John}},
  \bibinfo{author}{\bibfnamefont{M.}~\bibnamefont{Berritta}},
  \bibinfo{author}{\bibfnamefont{D.}~\bibnamefont{Hinzke}},
  \bibinfo{author}{\bibfnamefont{C.}~\bibnamefont{M{\"u}ller}},
  \bibinfo{author}{\bibfnamefont{T.}~\bibnamefont{Santos}},
  \bibinfo{author}{\bibfnamefont{H.}~\bibnamefont{Ulrichs}},
  \bibinfo{author}{\bibfnamefont{P.}~\bibnamefont{Nieves}},
  \bibinfo{author}{\bibfnamefont{J.}~\bibnamefont{Walowski}},
  \bibinfo{author}{\bibfnamefont{R.}~\bibnamefont{Mondal}},
  \bibinfo{author}{\bibfnamefont{O.}~\bibnamefont{Chubykalo-Fesenko}},
  \bibnamefont{et~al.}, \bibinfo{journal}{Sci. Rep.}
  \textbf{\bibinfo{volume}{7}}, \bibinfo{pages}{4114} (\bibinfo{year}{2017}),
  \urlprefix\url{http://adsabs.harvard.edu/cgi-bin/nph-data_query?bibcode=2017NatSR...7.4114J&link_type=EJOURNAL}.

\bibitem[{\citenamefont{Illg et~al.}(2013)\citenamefont{Illg, Haag, and
  F{\"a}hnle}}]{Illg:2013jp}
\bibinfo{author}{\bibfnamefont{C.}~\bibnamefont{Illg}},
  \bibinfo{author}{\bibfnamefont{M.}~\bibnamefont{Haag}}, \bibnamefont{and}
  \bibinfo{author}{\bibfnamefont{M.}~\bibnamefont{F{\"a}hnle}},
  \bibinfo{journal}{Phys. Rev. B} \textbf{\bibinfo{volume}{88}},
  \bibinfo{pages}{214404} (\bibinfo{year}{2013}),
  \urlprefix\url{https://link.aps.org/doi/10.1103/PhysRevB.88.214404}.

\bibitem[{\citenamefont{Evans et~al.}(2015)\citenamefont{Evans, Atxitia, and
  Chantrell}}]{Evans:2015ffa}
\bibinfo{author}{\bibfnamefont{R.~F.~L.} \bibnamefont{Evans}},
  \bibinfo{author}{\bibfnamefont{U.}~\bibnamefont{Atxitia}}, \bibnamefont{and}
  \bibinfo{author}{\bibfnamefont{R.~W.} \bibnamefont{Chantrell}},
  \bibinfo{journal}{Phys. Rev. B} \textbf{\bibinfo{volume}{91}},
  \bibinfo{pages}{144425} (\bibinfo{year}{2015}),
  \urlprefix\url{https://link.aps.org/doi/10.1103/PhysRevB.91.144425}.

\bibitem[{\citenamefont{Hinzke et~al.}(2015)\citenamefont{Hinzke, Atxitia,
  Carva, Nieves, Chubykalo-Fesenko, Oppeneer, and Nowak}}]{Hinzke:2015kra}
\bibinfo{author}{\bibfnamefont{D.}~\bibnamefont{Hinzke}},
  \bibinfo{author}{\bibfnamefont{U.}~\bibnamefont{Atxitia}},
  \bibinfo{author}{\bibfnamefont{K.}~\bibnamefont{Carva}},
  \bibinfo{author}{\bibfnamefont{P.}~\bibnamefont{Nieves}},
  \bibinfo{author}{\bibfnamefont{O.}~\bibnamefont{Chubykalo-Fesenko}},
  \bibinfo{author}{\bibfnamefont{P.~M.} \bibnamefont{Oppeneer}},
  \bibnamefont{and} \bibinfo{author}{\bibfnamefont{U.}~\bibnamefont{Nowak}},
  \bibinfo{journal}{Phys. Rev. B} \textbf{\bibinfo{volume}{92}},
  \bibinfo{pages}{259} (\bibinfo{year}{2015}),
  \urlprefix\url{https://link.aps.org/doi/10.1103/PhysRevB.92.054412}.

\bibitem[{\citenamefont{Atxitia et~al.}(2010)\citenamefont{Atxitia, Hinzke,
  Chubykalo-Fesenko, Nowak, Kachkachi, Mryasov, Evans, and
  Chantrell}}]{Atxitia:2010fu}
\bibinfo{author}{\bibfnamefont{U.}~\bibnamefont{Atxitia}},
  \bibinfo{author}{\bibfnamefont{D.}~\bibnamefont{Hinzke}},
  \bibinfo{author}{\bibfnamefont{O.}~\bibnamefont{Chubykalo-Fesenko}},
  \bibinfo{author}{\bibfnamefont{U.}~\bibnamefont{Nowak}},
  \bibinfo{author}{\bibfnamefont{H.}~\bibnamefont{Kachkachi}},
  \bibinfo{author}{\bibfnamefont{O.~N.} \bibnamefont{Mryasov}},
  \bibinfo{author}{\bibfnamefont{R.~F.} \bibnamefont{Evans}}, \bibnamefont{and}
  \bibinfo{author}{\bibfnamefont{R.~W.} \bibnamefont{Chantrell}},
  \bibinfo{journal}{Phys. Rev. B} \textbf{\bibinfo{volume}{82}},
  \bibinfo{pages}{134440} (\bibinfo{year}{2010}),
  \urlprefix\url{https://link.aps.org/doi/10.1103/PhysRevB.82.134440}.

\bibitem[{\citenamefont{Palmstr{\o}m}(2016)}]{Palmstrom:2016ju}
\bibinfo{author}{\bibfnamefont{C.~J.} \bibnamefont{Palmstr{\o}m}},
  \bibinfo{journal}{Progress in Crystal Growth and Characterization of
  Materials} \textbf{\bibinfo{volume}{62}}, \bibinfo{pages}{371}
  (\bibinfo{year}{2016}),
  \urlprefix\url{http://linkinghub.elsevier.com/retrieve/pii/S0960897416300237}.

\end{thebibliography}

\end{document}